\documentclass[twocolumn,]{aastex631}

\usepackage[T1]{fontenc}
\usepackage{graphicx}	% Including figure files
\usepackage{amsmath}	% Advanced maths commands
\usepackage{mathrsfs} % include special English Character
\usepackage{xcolor}
\usepackage{booktabs}

\newcommand{\revise}[1]{\color{black}{#1}\color{black}}
\newcommand{\ud}{\mathrm{d}}
\newcommand{\hbt}{\textsc{hbt+}}
\newcommand{\msunh}{\mathrm{M}_\odot\;h^{-1}}
\newcommand{\LCDM}{$\Lambda$CDM }

\begin{document}

\title{Self-similar decomposition of the hierarchical merger tree of dark matter halos}
\shorttitle{self-similar decomposition of merger tree}
\shortauthors{Jiang et al.}

\author{Wenkang Jiang}
\affiliation{Department of Astronomy, School of Physics and Astronomy, Shanghai Jiao Tong University, Shanghai, 200240, China}
\affiliation{State Key Laboratory of Dark Matter Physics, School of Physics and 
Astronomy, Shanghai Jiao Tong University, Shanghai 200240, China}

\author[0000-0002-8010-6715]{Jiaxin Han}
\affiliation{Department of Astronomy, School of Physics and Astronomy, Shanghai Jiao Tong University, Shanghai, 200240, China}
\affiliation{State Key Laboratory of Dark Matter Physics, School of Physics and 
Astronomy, Shanghai Jiao Tong University, Shanghai 200240, China}

\author[0000-0003-0296-0841]{Fuyu Dong}
\affiliation{South-Western Institute for Astronomy Research, Yunnan University, Kunming 650500, China}

\author[0000-0003-0303-4188]{Feihong He}
\affiliation{Department of Astronomy, School of Physics and Astronomy, Shanghai Jiao Tong University, Shanghai, 200240, China}
\affiliation{State Key Laboratory of Dark Matter Physics, School of Physics and 
Astronomy, Shanghai Jiao Tong University, Shanghai 200240, China}

\correspondingauthor{Jiaxin Han}
\email{eventually@sjtu.edu.cn(WK), jiaxin.han@sjtu.edu.cn(JH)}



\begin{abstract}
\revise{In the $\Lambda$CDM universe, structure formation is generally not a self-similar process, while some self-similarity remains in certain statistics which can greatly simplify our description and understanding of the cosmic structures.} In this work, we show that the merger tree of dark matter halos is approximately self-similar by investigating the universality of the subhalo peak mass function (PMF) describing the mass distribution of progenitor halos. Using a set of cosmological simulations and identifying subhalos of different merger levels with \hbt, we verify that the level-1 subhalo PMF is close to universal across halo mass, redshift, and cosmology. This approximate self-similarity allows us to analytically derive the subhalo PMF for subhalos accreted at any level (i.e., for sub-sub...halos) through self-convolutions of the level-1 PMF, and the resulting model shows good agreement with simulation measurements. We further derive a number of analytical properties on the hierarchical origin of subhalos. We show that higher-level subhalos dominate at progressively lower peak mass in the PMF and are more likely to originate from major mergers than lower-level ones. At a given merger mass ratio, the subhalo accretion rates at each level track the growth rate of the host halo. At a fixed final mass ratio, however, subhalos of higher-level, higher-mass-ratio and in more massive haloes tend to be accreted more recently. \revise{Matching subhalo peak mass to galaxy mass, these results have direct implications on the hierarchical origin of satellite galaxies.}
\end{abstract}



\section{Introduction} 
\label{sec:intro}
\revise{%The establishment of a universal halo merger rate and PMF can also have more general implications for our understanding of the structure formation process, as such laws can be considered as a manifestation of self-similarity. 
The seek for self-similarity used to be a fundamental topic in the study of structure formation~\citep[e.g.,][]{Ps74,Peebles74,Peebles85,Miyoshi75,Davis77,Gott78,Efstathiou79,Efstathiou81,Fry80,EF88}. As gravity is scale-free, when further coupled with a self-similar initial condition, then the process and outcome of structure formation can be expected to be self-similar, so that the distribution and evolution of structures of different sizes follow the same laws when rescaled by their sizes. For example, the collapse of a dark matter halo is self-similar for power-law initial conditions in the Einstein de Sitter (EdS) universe~\citep{FG84}. The formation of large scale structure is self-similar in the EdS universe with power-law initial conditions~\citep[e.g.,][]{EF88}. 

In the \LCDM universe, however, structure formation is not exactly self-similar, as the initial power spectrum is scale dependent and dark energy introduces another scale. Despite this, some (approximate) self-similarity are still found in many statistics of the \LCDM universe. The identification of these laws can greatly simplify our description of structure formation, while also bringing crucial insights about the physical processes. On one hand, accurate universal models of the clustering of matter in the nonlinear regime have been developed~\citep{HKLM,PD94,Smith03}, which started from analytical insights brought by self-similarity~\citep[][]{Peebles74,Peebles80, Davis77}. On the other hand, the evolution of the cosmic structure described in terms of dark matter halos has also revealed much self-similarity. For example, the Extended Press Schechter theory predicts that the halo mass function, as well as its time evolution, is universal when formulated with abstract variables such as the peak height, $\nu$, and collapse barrier, $\omega$, which can be regarded as a manifestation of self-similarity in this particular parameter space~\citep[e.g.,][]{Ps74, Lacey93, Lacey94, elipsoidal_sheth,Moreno08,time-similarity1}. %  the conditional progenitor mass function of dark matter halos is self-similar over time, when expressing the time variable with the collapse barrier in the Extended Press Schechter theory~\citep{ND08, Genel08, Fakhouri08, time-similarity1}. 
However, in terms of intuitive properties such as the mass of halos, self-similarity becomes rarer to find, due to the complexities introduced by scale dependence in the initial condition and structure formation. A notable exception is found in the distribution of subhalos. The subhalo mass function, expressed as a function of the ratio between the subhalo mass and the host halo mass, has long been shown to be largely self-similar~\citep[e.g.,][]{Gao04b,Stewart08,Giocoli08,vdb14}. Recently, \citet{Daneng23} also observed self-similarity in the graph showing the spatial hierarchy of the surviving subhalos in the Fire and Fire2 simulation. 

The origin of the approximate self-similarity in the subhalo distribution has to be traced back to how halos merge and their tidal evolutions thereafter~\citep[e.g.,][]{Han16,He24}. The distribution of the merger events can be studied in two equivalent approaches, by either studying the merger rate of halos, or by studying the distribution of progenitor masses. It thus can be generally expected that some self-similarity could be found in these two statistics.} %In this work, we expand upon previous findings about the self-similarity in subhalo distribution to show that the halo merger tree itself can be well approximated as a self-similar structure, by studying the distribution of the peak mass of accreted subhalos, i.e., the Peak Mass Function (PMF).} %It can be naively expected that the mergers among halos also proceed in a scale-free or self-similar manner, with smaller systems being scaled down versions of larger ones. However, considering the shape of the power spectrum, it is not expected that such self-similarity can be preserved down to a small halo mass range.} 

\revise{Along the first line, it has taken some efforts before self-similarity is identified in numerical simulations.} % Many works have attempted to establish empirical laws for the merger rate in numerical simulations. 
%\rev{Indeed, studies of merger rates in numerical simulations}
It has been generally found that the merger rate takes on a power law form at the low mass ratio end with an exponential decay at the high ratio end~\citep{Fakhouri08, Fakhouri09, Fakhouri10a, merger_r3, merger_r4}. However, different works have reported different dependencies of the merger rate on halo mass and redshift. For example, \citet{Fakhouri08} found a weak dependence on the progenitor mass (\revise{$\propto M_{\rm h}^{0.08}$}) and redshift (\revise{$\propto \frac{d\delta_c}{dz}^{0.3}$}) using the Millennium simulation~\citet{Springel01} up to $z=16$. On the other hand,~\citet[]{merger_r3, merger_r4} found a more rapid increase with respect to progenitor mass and redshift for $z$ smaller than 1 (\revise{$\propto M_{\rm h}^{0.15} (\frac{d\delta_c}{dz})^2$;$\propto M_{\rm h}^{0.12} (\frac{d\delta_c}{dz})$ separately}), and~\citet{merger_r6} reported a slightly higher major merger rate with a slower increase of the merger rate across redshift. 

The differences among various works may arise from both complexities in the merger tree construction and differences in the background cosmologies. For the former, it is widely accepted that the performances of halo merger trees can be sensitive to the choice of halo finders and the methods to handle pathological cases in merger trees~\citep[][]{sussing_merger1,sussing_merger2,sussing_merger3,sussing_merger4}. For example,~\citet{Fakhouri08} showed that different algorithms for addressing the 'fragmentation' problem had a pronounced effect on the merger rate. For the latter, different $\Omega_{\mathrm{M_0}}$ and $\sigma_8$ lead to different growth rates and variances of the smoothed density field separately, thus affecting the halo merger rate globally~\citep[][]{Lacey94,merger_mah_coupled1}.

In \revise{a more recent work}, ~\citet{Dong22} demonstrated that if the merger rate per halo was normalized by the logarithmic mass growth rate of host halos, the newly defined specific merger rate, expressed as $\mathrm{d}N_{\mathrm{merge}} / N_{\mathrm{halo}}/ \mathrm{d}\xi / \mathrm{dln}M_h $, exhibits a remarkable universality across the redshift range of $[0,5]$ and the halo mass range of $[\mathrm{10^{12},10^{14}}]\mathrm{M_{\odot}}$. This universality can also be maintained under different background cosmologies with varying values of $\Omega_{\mathrm{M_0}}$ and $\sigma_{8}$. ~\citet{Dong22} also pointed out that the specific merger rate is equivalent to the derivative of the Peak Mass Function (PMF) — an alternative approach in describing the distribution of the merging population, which specifies the number of progenitor halos of a given infall mass accumulated through all previous redshifts. Thus, it is expected that the PMF should exhibit the same universality as the merger rate.

\revise{The PMF has been extensively studied using various approaches, including numerical simulations~\citep[e.g.,][]{Giocoli08,Stewart08,Li09,Han18,Xie15,Chua}, Monte Carlo merger trees~\citep[e.g.,][]{vdb05,vdb14}, and theoretical models~\citep[e.g.,][]{Salvador_accsub,Giocoli_08c,Yang11}. These studies have consistently shown that the PMF can be well described by a Schechter function. While some studies suggest the PMF is universal across halo mass ranges~\citep[e.g.,][]{Giocoli08,Li09,Han18}, others, such as \citet{Xie15}, report a clear mass dependence in the Aquarius and Phoenix simulations, particularly when excluding `re-accreted' and `ejected' subhalos from the unevolved population. In contrast, the redshift dependence of the PMF has been less explored, with only a few studies examining its universality in cosmological simulations~\citep[e.g.,][]{Li09,Giocoli_conf}. Furthermore, the influence of varying cosmological parameters on the PMF remains largely unstudied in simulations. Therefore, it remains to be investigated whether and to what extent the PMF can be modeled as universal, and what subhalo mass and membership definitions lead to the maximum amount of universality %critical to determine the optimal definition of subhalo mass and population that preserves consistency 
across halo mass, redshift, and cosmology, in correspondence to the universality of the specific halo merger rate.}

\revise{In this study, we adopt HBT+ subhalo finder~\citep{HBT,Han18}, a time-domain subhalo finder known for its robustness and consistency, to identify subhalos and build the merger tree. We systematically examine the universality of the PMF in simulations, addressing several significant issues that arise in this context. Firstly, a definition for the membership of the subhalo, as well as a definition for the peak mass, are required when constructing the PMF. Second, depending on the halo finder and mass definition, the masses of the subhalos may depend on the resolution of the simulation. To this end, we investigate different combinations of subhalo mass and membership definitions, and identify the one that best preserves the universality of the PMF. } A number of N-body simulations with different cosmologies and resolutions are employed to investigate the physical and numerical dependence of the results.

A key prediction of the hierarchical structure formation paradigm is that halos merge hierarchically, leading to a tree-like structure describing the merger histories of halos, and a corresponding hierarchy of subhalos with different levels. 
Previous studies on the merger rate of halos mostly focus on direct mergers among halos, corresponding to the progenitors of first-level subhalos~\citep[e.g.,][]{merger_r2,Fakhouri09,Fakhouri10a,merger_r4,merger_r6,Dong22}. In terms of the PMF, subhalo progenitors without selection in level, i.e., at all levels as a whole, are also frequently studied~\citep[e.g.,][]{Li09, Xie15,Chua, Han18}. The use of \hbt allows us to obtain the PMFs for subhalos at different levels separately. If the PMF of the first level is universal across mass and redshift, an immediate implication is that the PMFs of all levels can be derived through convolutions of the first level function. The resulting universal PMFs are then a direct manifestation of the self-similarity of the halo merger tree, such that the tree truncated at any branch has the same branching properties as the root branch. In the second part of this study, we study how well different levels of the PMF can be modeled this way. The derived model is further used to predict a few population properties of subhalo progenitors in redshift, mass, and level.

As subhalos correspond to the hosts of satellite galaxies inside a larger galaxy system, the universal and analytical model for the multi-level PMFs obtained here may find applications in a wide range of studies related to subhalos, including the structure and origin of stellar streams and the stellar halo~\citep[e.g.,][]{Stream1,Stream2,Stream3,Stream4,HeJ24,Tan24}, the formation of globular clusters~\citep[e.g.,][]{GC1,GC2,GC3}, the cosmological formation and evolution of galaxies~\citep[e.g.,][]{Springel01,Yang12}, and the particle nature of dark matter~\cite[e.g.,][]{identity1,identity2,feihong,Springel08dm,Gao11,Han12,Han16}.

This paper is organized as follows. In Section~\ref{sim_intro}, we introduce the simulation data. Section~\ref{sec:uni-check} presents the model used to construct PMFs of different levels, along with an investigation into the universality of these PMFs and the performance of the model. Subsequently, Section~\ref{sec:hierarchical_origin} discusses several theoretical implications of the universal hierarchical PMFs developed in the previous section. In Section~\ref{sec:discussion}, we examine systematics in the PMFs resulting from different mass and membership combinations, as well as the excessive peak mass caused by insufficient particle resolution. Finally, we summarize and conclude in Section~\ref{sec:conclusion}.

In the context of studying the PMF, the peak mass of a subhalo represents the mass of its progenitor, and the statistical properties derived from the PMF actually describe the properties of the progenitors. We assume it is understood that when we talk about subhalos in this context, we refer to the progenitor population.

\section{Simulation and subhalo catalog} \label{sim_intro}
\begin{table*}[htb!]
    \centering
    \caption{Cosmological parameters setting for different simulations.}
    \label{tab:cos_param}
    
    \begin{tabular}{cccccccc} 
    \hline
    \hline
    & $\Omega_{\mathrm{M_0}}$ & $\Omega_{\mathrm{\Lambda_0}}$ & $\sigma_{8}$ & $n_{\mathrm{s}}$  & $N_{\mathrm{p}}$ & $m_{\rm p}({\rm M_{\odot}}\;h^{-1})$ & $L_{\rm box}(\mathrm{Mpc}\;h^{-1})$\\
	\tableline
    \textbf{L600}  &  0.268 & 0.732 & 0.863 & 0.963  &$3072^3$   & $5.54\times10^{8}$ &  600\\
    \tableline
    \textbf{S02}   &  0.2   & 0.8   & 0.863   & 0.963   & $512^{3}$ & $4.14\times10^{8}$ & 100\\
    \tableline
    \textbf{S04}   &  0.4   & 0.6   & 0.863   & 0.963   & $512^{3}$ & $8.27\times10^{8}$ & 100 \\
    \tableline
    \textbf{S09}   &  0.9   & 0.1   & 0.863   & 0.963   & $512^{3}$ & $1.86\times10^{9}$ & 100\\ 
    \tableline
    \textbf{Sg87}   &  0.268   & 0.732   & 0.87   & 0.963   & $512^{3}$ & $5.54\times10^{8}$ &  100\\
    \tableline
    \textbf{Sg80}   &  0.268   & 0.732   & 0.80   & 0.963   & $512^{3}$ & $5.54\times10^{8}$ &  100\\
    \tableline
    \textbf{Sg50}   &  0.268   & 0.732   & 0.50   & 0.963   & $512^{3}$ & $5.54\times10^{8}$ &  100\\
    \tableline
    \textbf{Sg30}   &  0.268   & 0.732   & 0.30   & 0.963  & $512^{3}$ & $5.54\times10^{8}$ &  100\\
    \tableline
    \textbf{NS01}   &  0.268   & 0.732   & 0.863   & 0.963 & $512^{3}$ & $5.54\times10^{8}$ &  100\\
   \tableline
    \textbf{NS02}   &  0.268   & 0.732   & 0.863   & 1.5  & $512^{3}$ & $5.54\times10^{8}$ &  100\\
   \tableline
    \textbf{NS03}   &  0.268   & 0.732   & 0.863   & 2.0   & $512^{3}$ & $5.54\times10^{8}$ &  100\\
    \tableline
    \textbf{Kanli}  &  0.3156 & 0.6844 & 0.81 & 0.967 & $2048^3$   & $1.02\times10^{7}$ & 100\\
    \tableline
	\end{tabular}
\end{table*}
Our work is based on five sets of $\Lambda$CDM N-body simulations shown in Table~\ref{tab:cos_param}. Among these simulations, L600 has the highest resolution and the largest box size, serving as the baseline simulation for drawing our main conclusions. It is one of the series of simulations in \textit{CosmicGrowth} ~\citep[][]{CosmicGrowth}. In addition to this baseline simulation, we conduct three subsequent sets of lower-resolution simulations using \textsc{Gadget4}~\citep{Gadget4}, with varying $\Omega_{\mathrm{M_0}}$, $\sigma_{8}$, and the primordial spectral index $n_{s}$. These simulations are used to study the influence of different cosmologies. Finally, we incorporate another CDM simulation with a lower particle mass. It is one of the series of simulations in \textit{Kanli}~\citep{feihong}. We can figure out the influence of the particle resolution by comparing the PMFs obtained from \textit{Kanli} and L600.
    
Halo catalogs in all the simulations are generated using the Friends of Friends algorithm~\citep[FoF,][]{FoF} with a linking length of $0.2$ times the mean particle separation, down to a lower mass limit of 20 particles. Meanwhile, we apply the Hierarchical Bound-Tracing ~\citep[\hbt,][]{HBT,Han18} algorithm to identify subhalos and build up the merger trees. \hbt is a unique subhalo finder that works in the time domain by tracking the evolution of halos to identify subhalos. It has demonstrated superb performance in generating highly consistent and physically accurate catalogs of subhalos and their evolution according to various tests~\citep[][]{Muldrew, HBT, Onions12, Srisawat, Behroozi15, Han18}, overcoming many systematics common to configuration space or even phase space subhalo finders.

We define the PMF as
\begin{equation}
    g(\mu)\equiv \frac{\mathrm{d}N_{\mathrm{sub}}}{\mathrm{d}\ln\mu},
\end{equation}
where 
\begin{equation}
    \mu\equiv m_{\rm peak}/M_h \label{eq:mu_def}
\end{equation} represents the mass ratio between the peak mass of a subhalo and the current mass of the host halo, and $N_{\mathrm{sub}}$ denotes the number of subhalos in the mass bin $[\mu, \mu + \mathrm{d} \mu]$ per host halo.  The level of universality in the PMF could differ depending on the exact definitions of the subhalo and host halo masses, as well as on how the subhalos are selected when counting their numbers for each halo. In \hbt, the mass of a subhalo is defined to be its self-bound mass, and we define the peak mass, $m_{\rm peak}$,  of a subhalo as the maximum bound mass reached during its evolutionary history. A significant fraction of subhalos could become unresolved (or ``disrupted") in simulations due to their high mass loss rates and the finite resolution of the simulation~\citep{Han16,He24}. Additionally, some subhalos could have merged with other subhalos or the host halo (defined as ``sink" in \hbt). It is necessary to include the progenitors of these disrupted or merged populations when computing the PMF, in order to account for all the subhalos that have ever merged into the host. In \hbt, a subhalo is tracked continuously with its most-bound particle once it becomes unresolved. This allows us to easily collect all the subhalos that have been accreted into a given host halo, no matter whether these subhalos have become unresolved or not. For the main part of this work, we count all the subhalos, including disrupted or merged ones, in each FoF group when computing the PMF, and define the mass of the host halo as the \revise{sum of the bound masses } of their member subhalos accordingly. This choice is different from many previous works, which use the virial mass for the host halo mass and only count subhalos inside the virial radius. We discuss the influence of different mass and membership definitions in Section~\ref{sec:discussion}. \revise{As the PMF at each redshift is determined independently from the subhalos belonging to the host halo at that time, any changes of subhalo membership including fly-by or ejection of subhalos are automatically accounted for.}

In this work, we focus not only on the PMF of the total population, $g_{\mathrm{t}}(\mu)$, but also on the PMFs of subhalos categorized by different accretion levels, $g_{\ell}(\mu)$. The accretion level describes the number of branches a subhalo has merged with throughout its previous evolution. For example, subhalos that have been directly accreted onto the main branches are characterized by $\ell =1$, while sub-subhalos, which are directly embedded within level-1 subhalos, are characterized by $\ell = 2$. In \hbt, the accretion level of each subhalo is recorded directly by the \texttt{Depth} parameter.

In \hbt, the masses of subhalos are defined to be exclusive, meaning that a subhalo does not include mass from its sub-subhalos. As a result, a subhalo resolved at the leaf of the hierarchy (i.e., of the highest subhalo level) will have its mass overestimated due to contributions of its unresolved sub-subhalos. It artificially inflates the peak mass of each subhalo, particularly for low-mass subhalos that are near the resolution limit, as these subhalos have almost no resolved substructure. Assuming the final subhalo mass function is also universal after being rescaled, we correct for this particle resolution effect as detailed in Section~\ref{sec:discussion}.

\section{The universality of the hierarchical PMFs} \label{sec:uni-check}
\subsection{Recurrence Relation of the PMFs} \label{sec:recursive-relation}
According to the hierarchical structure formation process, when a satellite halo merges into another halo to become a subhalo, it brings along its subhalos into the final host, increasing their levels by 1. The PMF of high-level subhalos is then produced by summing up the contribution of lower-level subhalos brought in by the halo mergers. This leads to a recurrence relation between the PMF of subhalos of two consecutive levels, $g_\ell$ and $g_{\ell-1}$, as
\begin{equation}
    \begin{aligned}
        g_\ell(\mu)=\int^{\infty}_{\mu'=0} g_{\ell-1}(\mu')g_{1}(\mu/\mu')\mathrm{d}\ln\mu'.
     \end{aligned}
    \label{eq:hier1_nobeta}
\end{equation}
In the above relation, $g_{\ell-1}$ specifies the distribution of level $\ell-1$ subhalos, while $g_1$ specifies the distribution of level $\ell$ subhalos embedded in level $\ell-1$ subhalos, with $\mu/\mu'$ specifying the peak mass ratio between the two levels.
In principle, these distributions could both depend on halo mass and redshift. However, if $g_1(\mu)$ is independent of host mass and redshift, then the above relation immediately implies that the $g_\ell$'s at all levels are also universal, and we can derive the PMF of any level by applying Equation~\eqref{eq:hier1_nobeta} recursively.

In the above derivation, however, a systematic effect we have ignored is that the peak mass ratio, $\mu/\mu'$, is not the exact mass ratio (Equation~\eqref{eq:mu_def}) required as input into the peak mass function. To correct for this difference in mass definition and achieve a more accurate description of high-level PMFs, we can introduce an effective correction factor, $\beta$, so that the recurrence relation becomes
\begin{equation}
    \begin{aligned}
        g_\ell(\mu)=\int^{\infty}_{\mu'=0} g_{\ell-1}(\mu')g_{1}(\beta\mu/\mu')\mathrm{d}\ln\mu'.
     \end{aligned}
    \label{eq:hier1}
\end{equation} This $\beta$ factor, can be interpreted as the average ratio between the peak mass and halo mass of a subhalo right before it turns into a satellite. 

Defining $x\equiv\ln\mu$, the above equation turns into a convolution equation and can be easily solved in Fourier space. If we define $h_\ell(x) \equiv g_\ell(e^{x})$ and the Fourier transformation of $h_\ell(x)$ as $\hat{h}_\ell(k)$, we have
\begin{equation}
\hat{h}_\ell(k)= (e^{ik{\rm ln}\beta})^{\ell-1} \hat{h}_1(k)^{\ell}.
\label{eq:h_n}
\end{equation} 
 The total PMF of all levels can be analytically linked to the level-1 PMF in Fourier space as \footnote{In practice, the PMF can only be resolved up to a finite level in a given simulation, so the summation in Equation~\eqref{eq:h_sum} should be limited to the same level range when fitting the simulation data.}
\begin{align}
    \hat{h}_t(k)&=\sum_{\ell=1}^{\mathrm{\infty}}\hat{h}_{\ell}(k)\label{eq:h_sum}\\
    &=\frac{\hat{h}_1(k)}{1-\hat{h}_1(k) e^{ik{\rm ln}\beta}}\label{eq:hier1-0}.
\end{align}
Accordingly, the level-1 PMF can be fit with the measured PMF of any level.

\subsection{Fitting the PMFs}
\begin{figure*}[htb!]
        \centering
        \includegraphics[width=0.48\textwidth,height=0.45\textwidth]{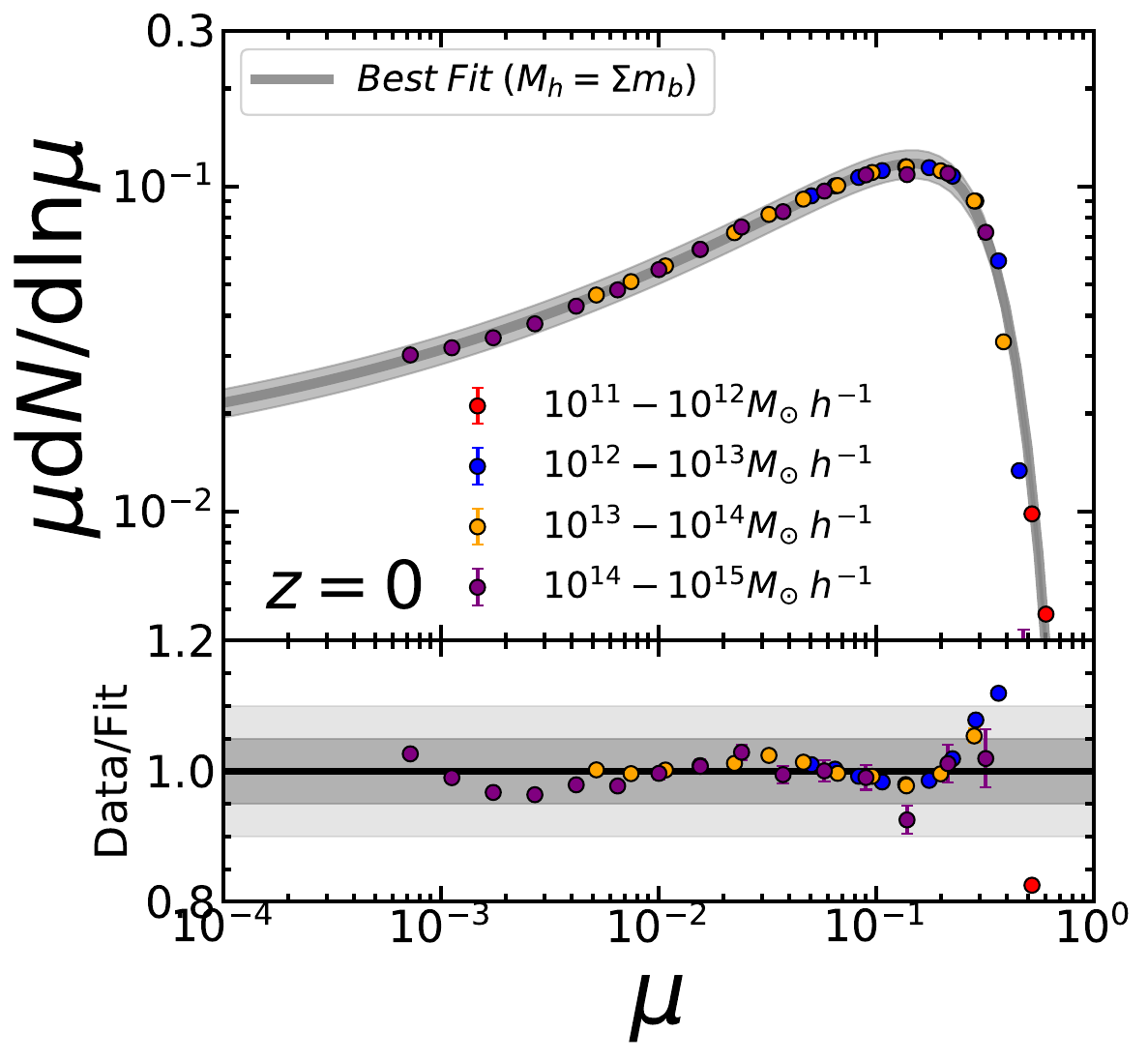}
        \includegraphics[width=0.48\textwidth,height=0.45\textwidth]{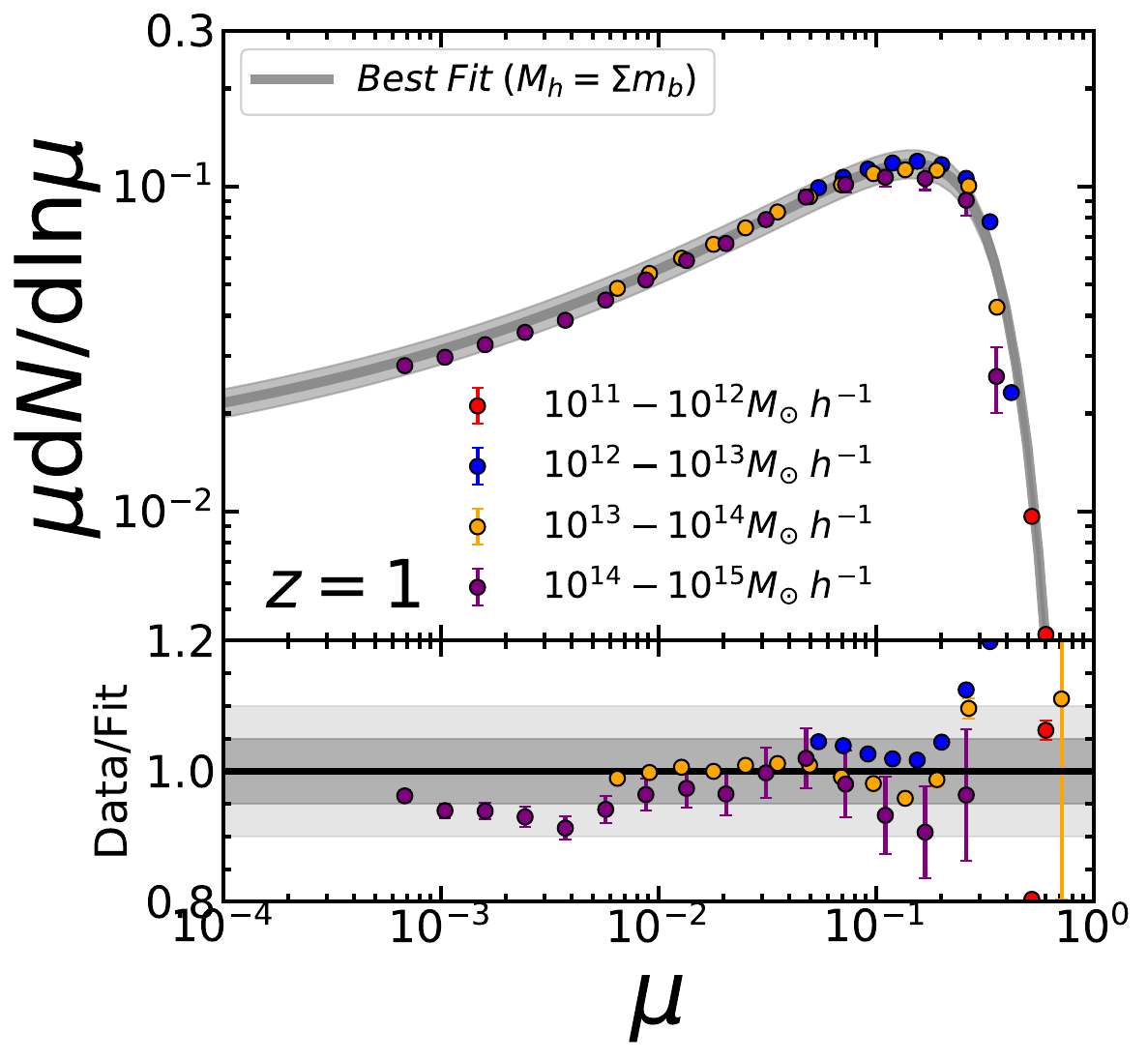}
        \includegraphics[width=0.48\textwidth,height=0.45\textwidth]{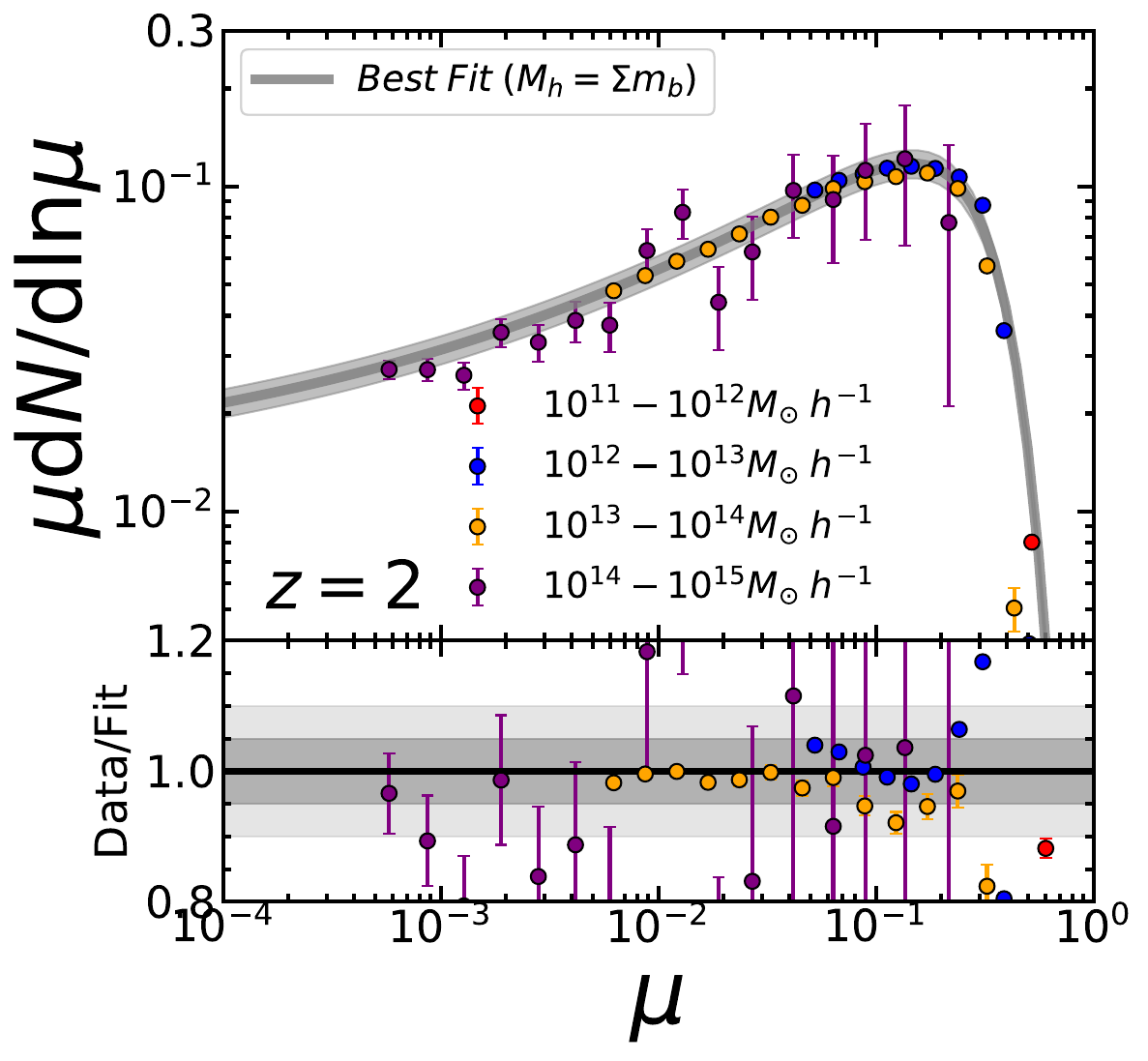}
        \includegraphics[width=0.48\textwidth,height=0.45\textwidth]{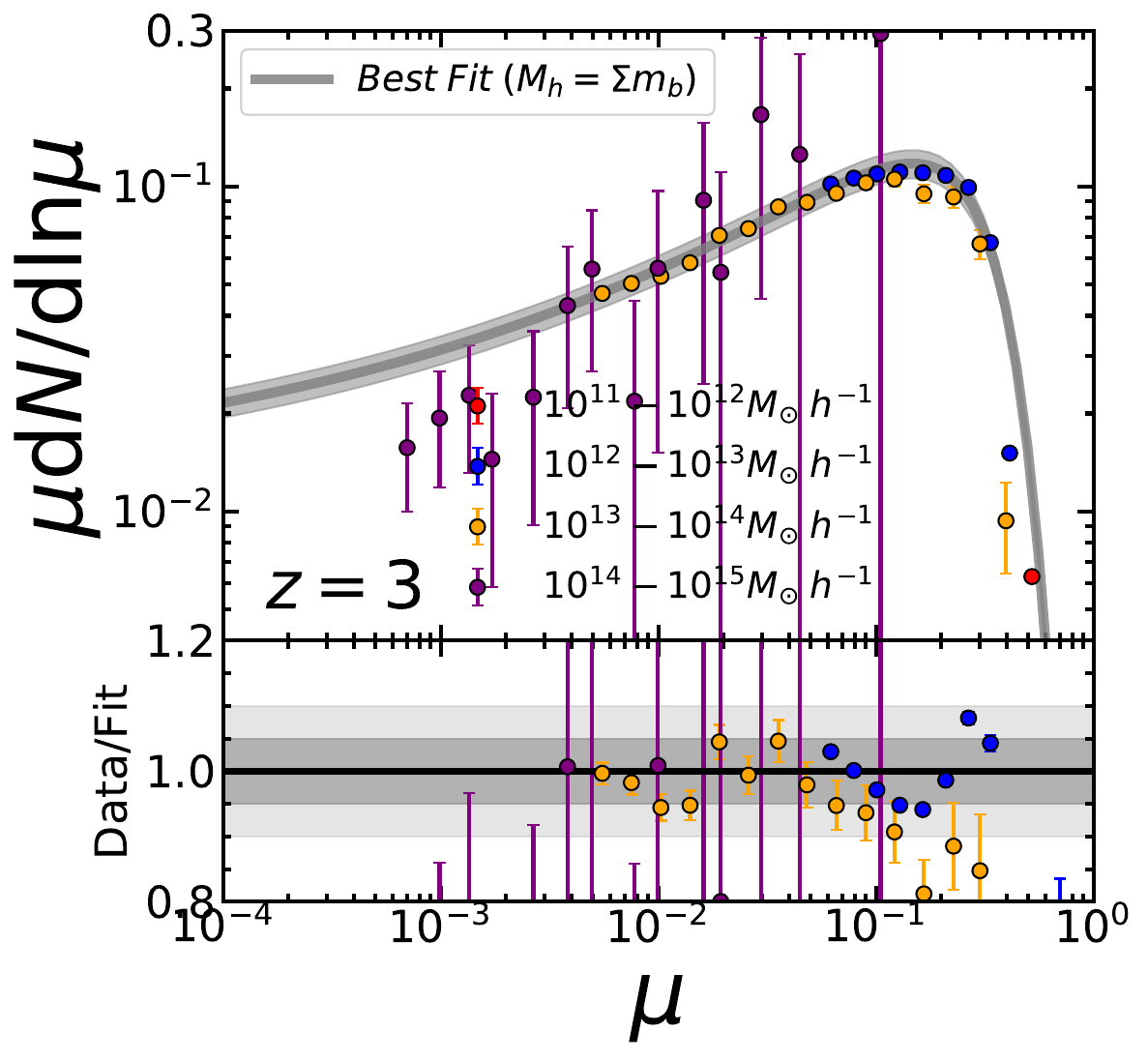}
\caption{The level-1 PMF across host halo mass and redshifts in the L600 simulation, as labeled. \revise{The data points show the average PMF over all halos in each specific mass bin and redshift. } The gray solid line shows the best-fit double Schechter function at $z=0$, reproduced in all the panels. The light gray shaded region shows a $\pm 10\%$ deviation, while the darker shaded region indicates a $\pm 5\%$ deviation from the fit. The bottom sub figure in each panel shows the ratio between the data and the fit. \revise{The errorbars show the uncertainty on the average PMF estimated from the Poisson noise in the total subhalo count in each bin.
}}
\label{fig:dependence_check}
\end{figure*} 

Figure~\ref{fig:dependence_check} shows the level-1 PMFs across various $M_h$ and $z$ in the baseline simulation L600. It can be seen that the measurements from different mass bins are very consistent with each other, and there is not any obvious redshift evolution. To assess the universality more qualitatively, we first fit the measurements at $z=0$ using a double Schechter function,
\begin{equation}
\label{eq:db}
\centering
    \begin{aligned}
        g_1(\mu)=\frac{ \mathrm{d}N}{\mathrm{dln}\mu}=(a_1\mu^{\alpha_1}+a_2\mu^{\alpha_2})e^{-c\mu^d},
    \end{aligned}
\end{equation} combining data points from all four mass bins. The best fitting parameters are
\begin{small}
    \begin{equation}
     (a_1,a_2,\alpha_1,\alpha_2,c,d)=(0.029,0.273,-0.94,-0.54,12.89,2.26).
    \label{eq:best_fit}
    \end{equation}    
\end{small}

This best-fitting model is overplotted in all the other panels, with a gray-shaded region representing the $\pm 10\%$ deviation. No significant variations (>10\%) are found between the data points and the fit, except for some data points with large Poisson errors.\revise{ We also conduct a brief analysis on the halo-to-halo variation on $g_1$ in Appendix~\ref{app:halo-to-halo_variation}. The results show that the scatter of the $g_1$ among different halos is close to or smaller than the Poisson noise in the region $\mu>10^{-2}$. For less massive subhalos $\mu< 10^{-2}$, the $g_1$ shows a mildly broader distribution than the Poisson distribution($1.0<\sigma/\sigma_{\rm Poisson}<1.6$), indicating the presence of a small amount of halo-to-halo scatter on top of the Poisson noise.}
\begin{figure*}[htb!]
    \centering
    \includegraphics[width=0.32\textwidth]{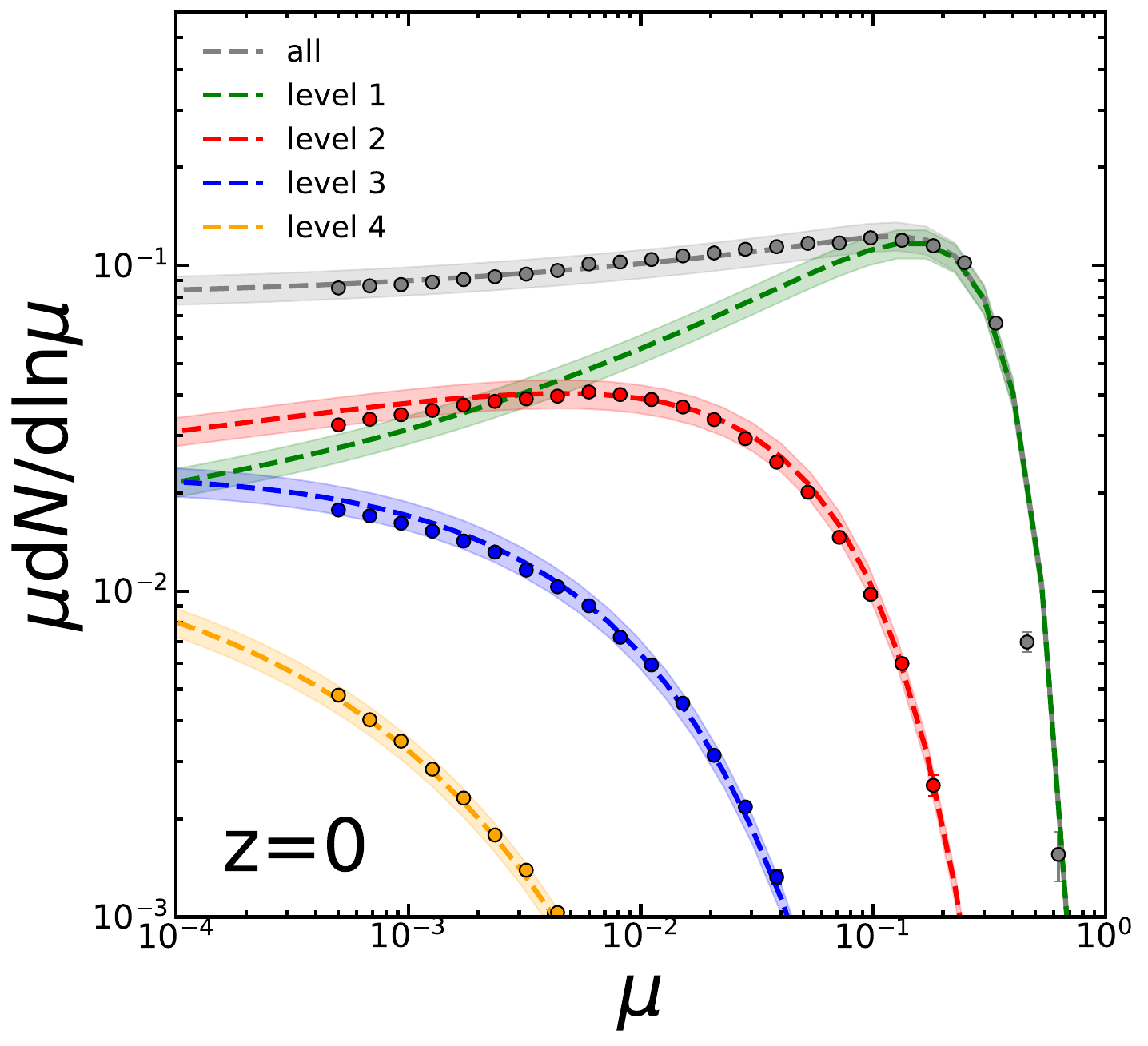}
    \includegraphics[width=0.32\textwidth]{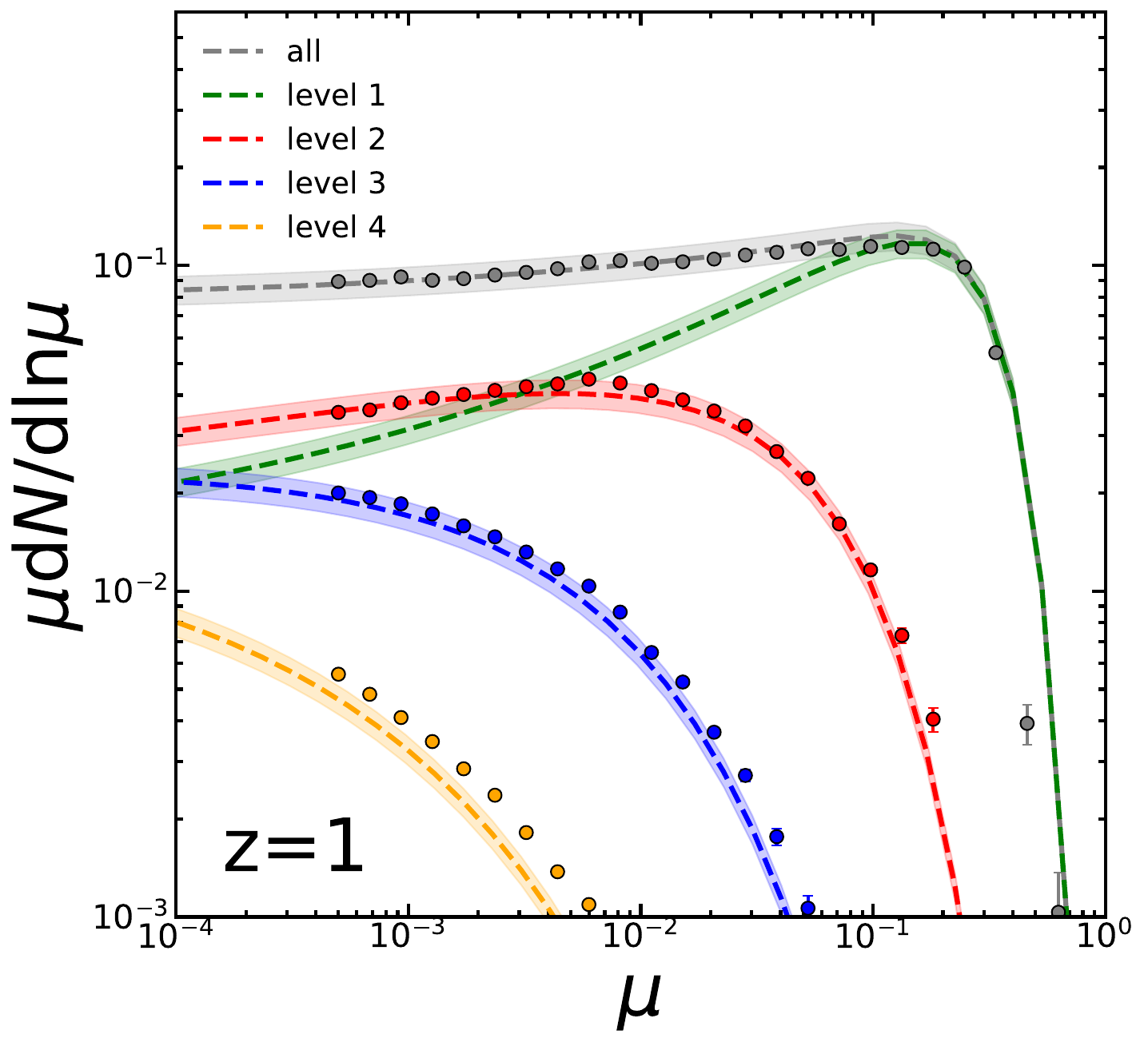}
    \includegraphics[width=0.32\textwidth]{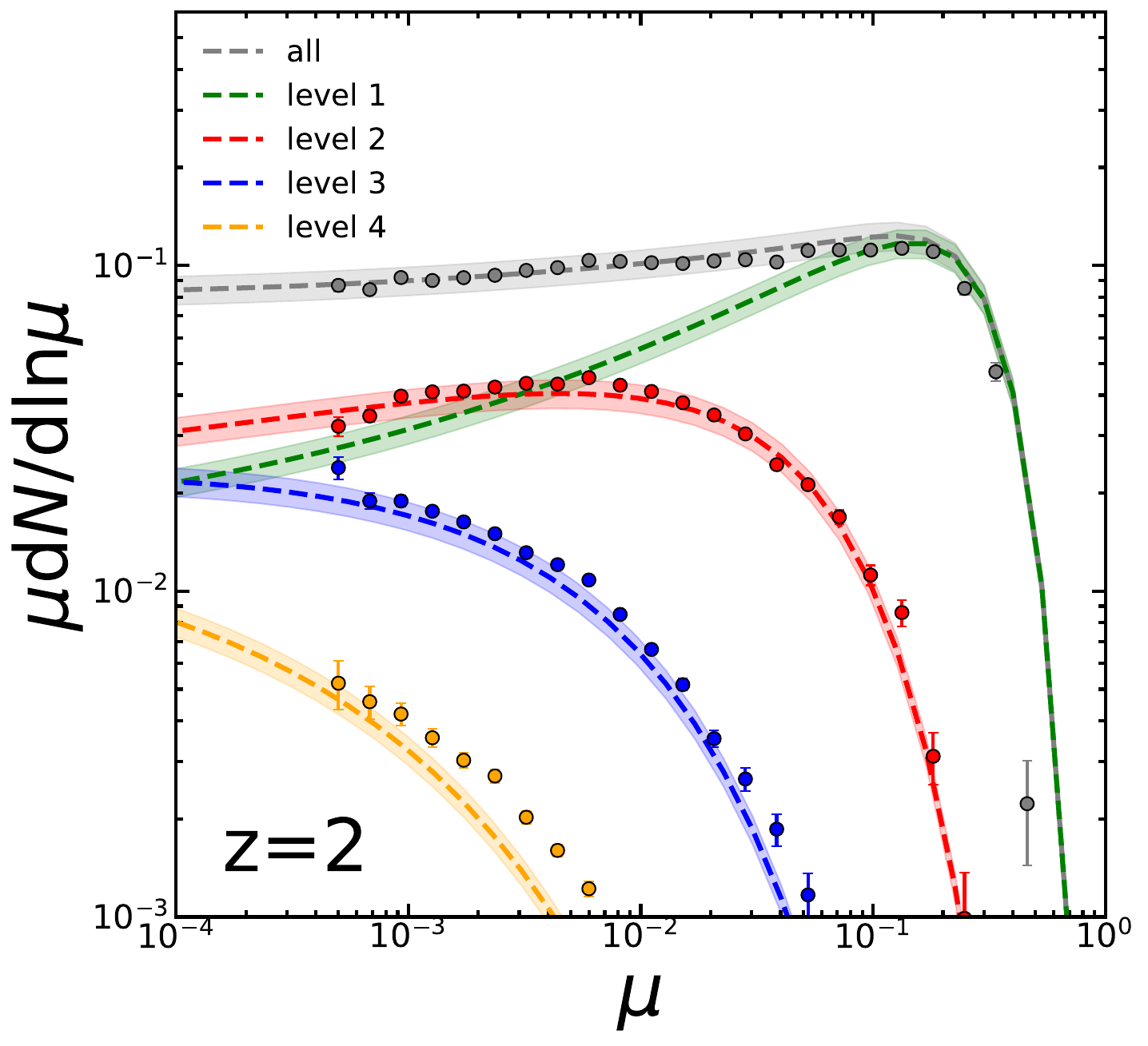}
                                                  
                         \caption{The PMFs of different levels and redshifts in the L600 simulation. Subhalos residing in different halo mass bins between $[10^{11},10^{15}]\mathrm{M_{\odot}}$ have been combined in the measurement. For clarity, we have not shown the measurements of the level-1 PMF. The dashed lines of different colors show the best-fit models at $z=0$ according to Equation~\eqref{eq:hier1}, reproduced in all three panels. Note the parameters controlling $g_1$ are fixed as in Fig.~\ref{fig:dependence_check}, while only the $\beta$ parameter is varied to fit the data at other levels. The shaded regions show the $\pm 10\%$. \revise{The errorbars show the uncertainty on the average PMF estimated from the Poisson noise in the total subhalo count in each bin.}
                         }
    \label{fig:convolution}
\end{figure*}

With a model of $g_1$ established, we further fit the PMFs of higher levels by varying the only remaining parameter $\beta$. At $z=0$, we obtain a best-fit parameter $\beta=0.726$.  As shown in Fig.~\ref{fig:convolution}, the model can well match the measurements at different levels simultaneously. The same model is also compared against measurements at other redshifts up to $z=2$, showing generally good agreement. At the highest level, there is a tendency that the PMF becomes underestimated at higher redshifts. This is indicative of a weak redshift evolution in the subhalo PMF. The deviation from the universal model is too small to be noticed in low-level PMFs but becomes amplified in high-level PMFs according to Equation~\eqref{eq:h_n}.
Nevertheless, the PMF at high redshift also becomes poorly resolved in the simulation. The weak redshift evolution will be examined in more detail in an upcoming work using the Jiutian simulation of a higher resolution~\cite{Jiutian}.
As the level-4 PMF contributes little to the total PMF, the universal model still works very well in predicting the total PMF.

\subsection{Dependence on Cosmology}
\begin{figure*}[htb!]
        \centering
        \includegraphics[width=0.32\textwidth]{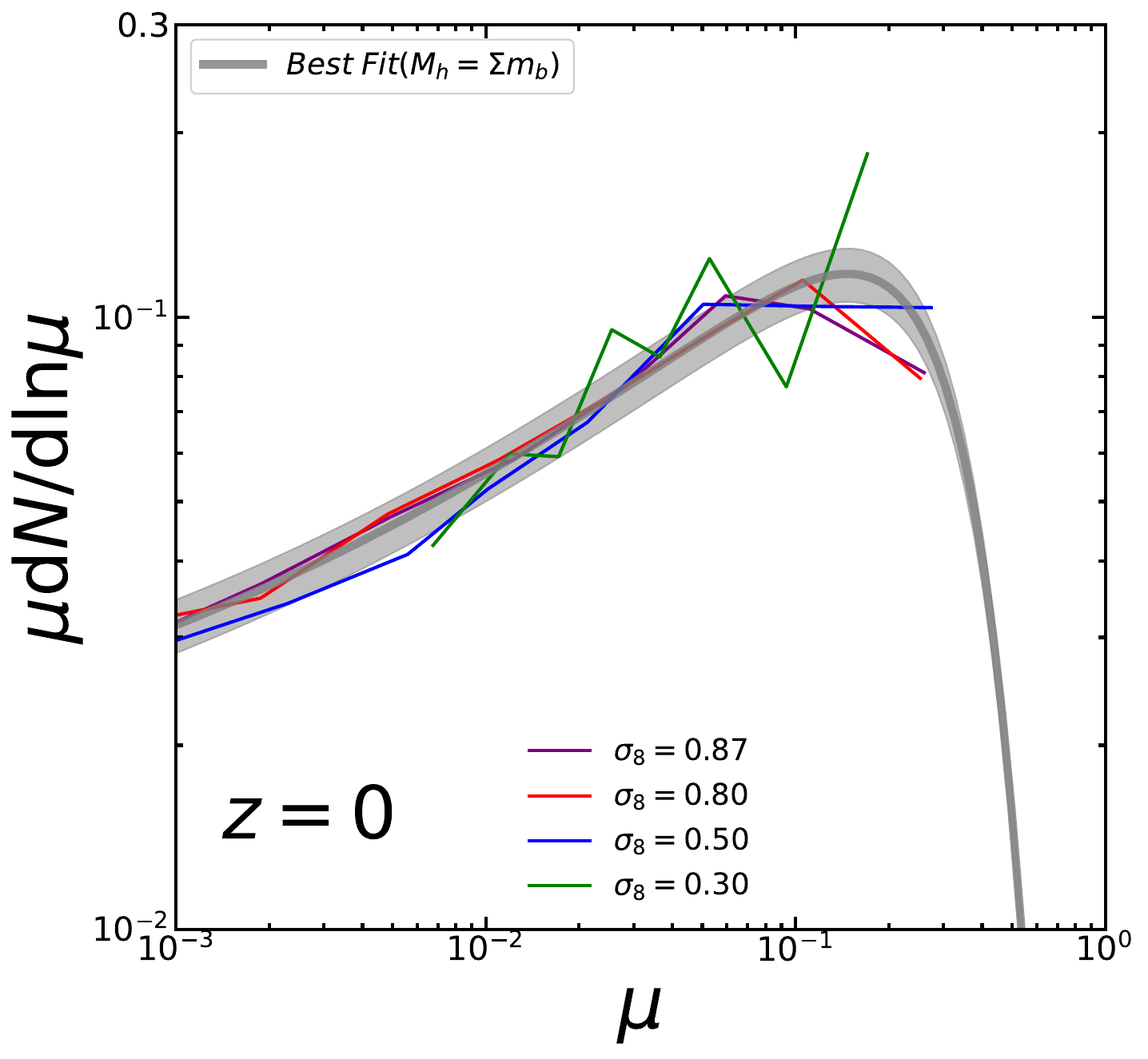}
        \includegraphics[width=0.32\textwidth]{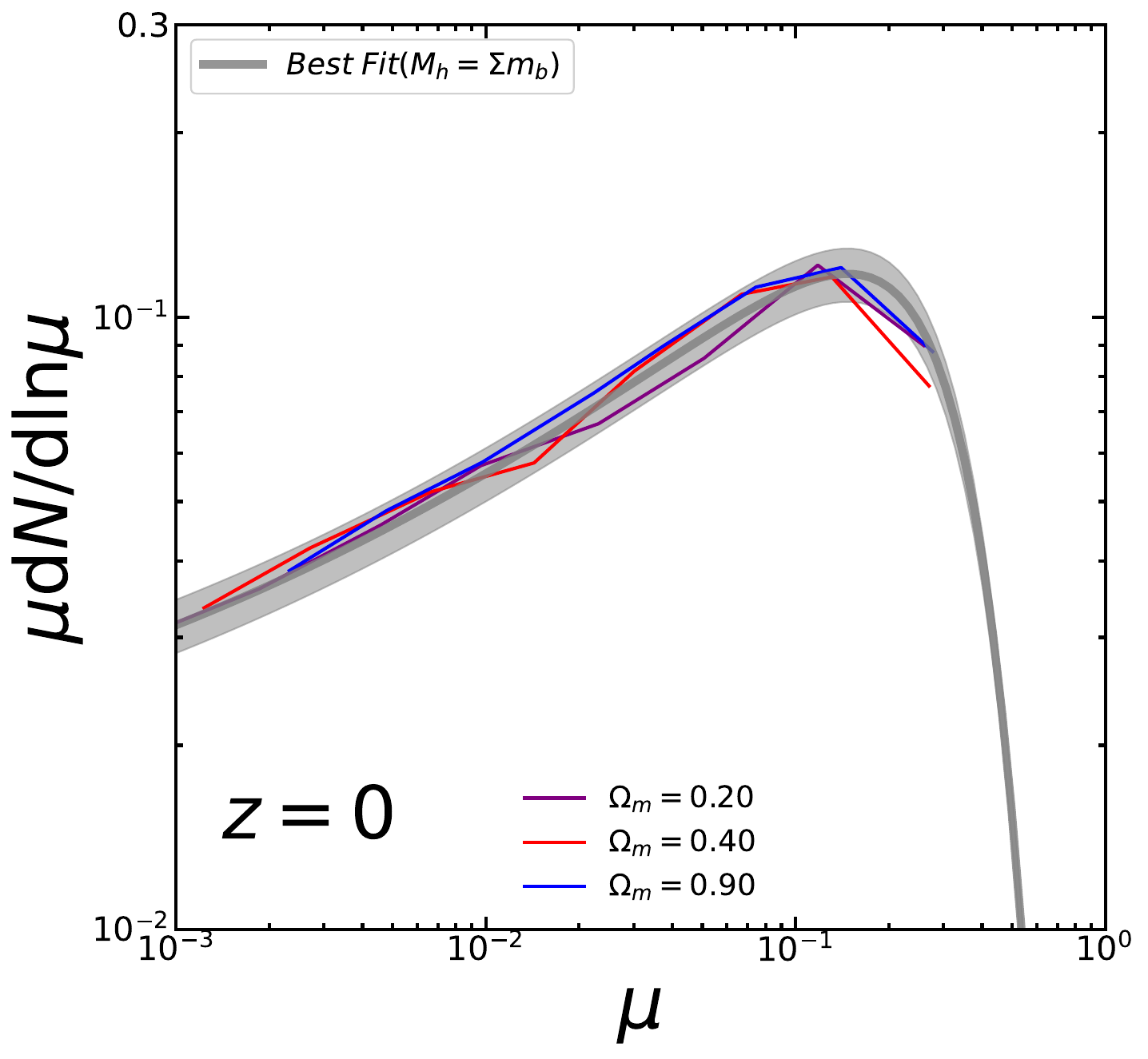}
        \includegraphics[width=0.32\textwidth]{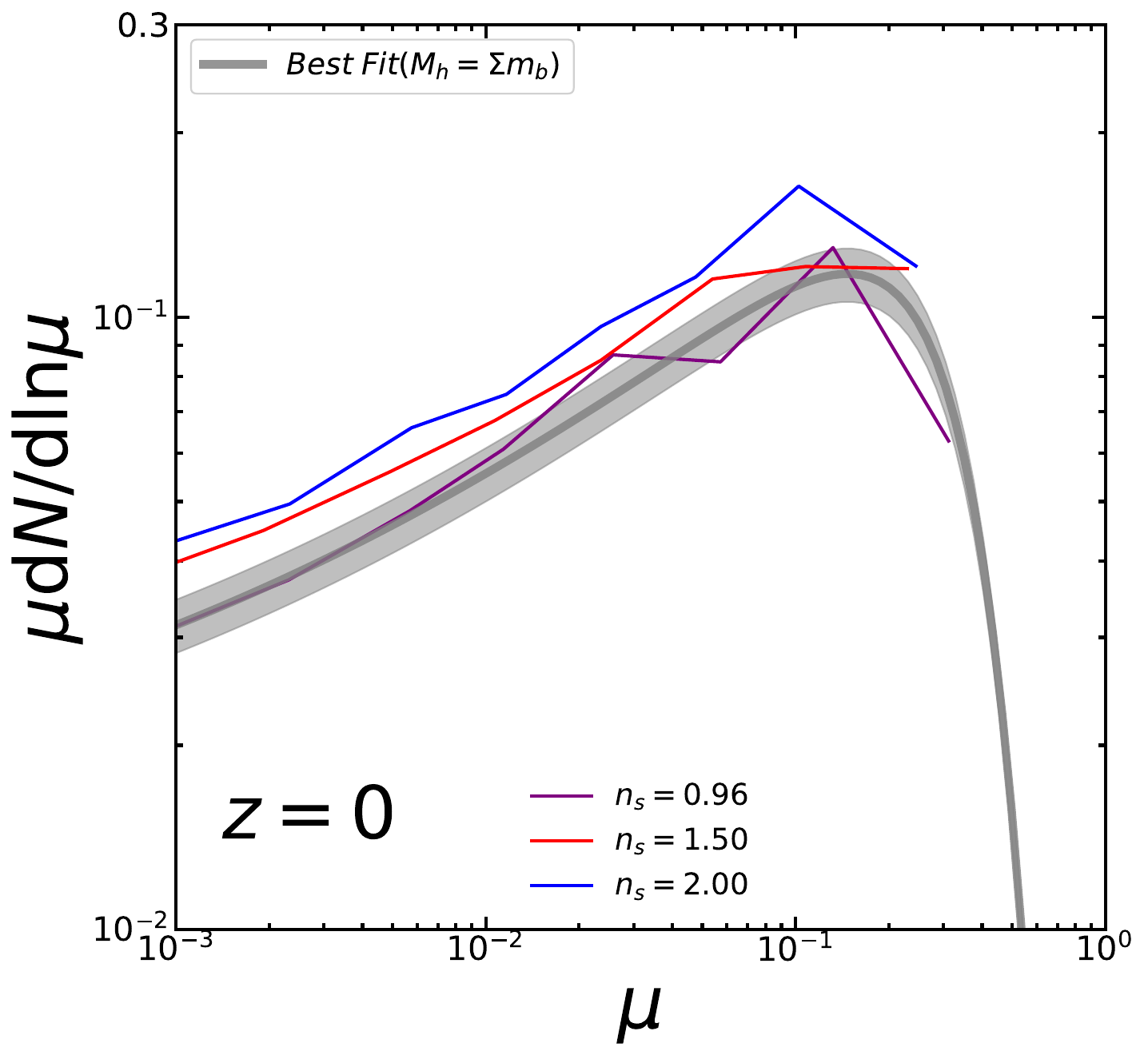}
\caption{ Level-1 PMF dependence on cosmological parameters $\sigma_8$, $\Omega_{\mathrm{M_0}}$, $n_\mathrm{s}$ measured at $z=0$ separately. The left panel shows the comparisons of $g_1$ under different $\sigma_8$ relative to the fiducial values. Their general trends trace the fiducial values as expected and scatters can be controlled within $\pm 10\%$ in most cases, despite some large fluctuations observed when extreme parameters are adopted. The middle panel shows the dependence on $\Omega_{\mathrm{M_0}}$, and their performances are similar to variations of $\sigma_8$. However, the diverse amplitudes of $g_1$ in the right panel show a strong correlation between accreted subhalos and primordial power spectrum index $n_\mathrm{s}$.} 
\label{fig:cosmo_depend}
\end{figure*}

So far we have only modeled the PMFs in a single simulation. To assess the universality of $g_1$ across different cosmologies, we show results from a set of simulations varying the parameters $\Omega_{\mathrm{M_0}}$, $\sigma_8$, and $n_\mathrm{s}$ in Figure~\ref{fig:cosmo_depend}. In each panel, the best-fitting model from the L600 simulation is shown with a gray solid line as a reference.

In the left and middle panels, no significant dependencies on $\Omega_{\mathrm{M_0}}$ and $\sigma_{8}$ are observed, and most of the measurements align well with the best-fitting result of L600 generally within 10 percent. Although some fluctuations are found in $g_1$ obtained from simulations with extreme parameters, these can be attributed to measurement uncertainties arising from limited sample sizes. For example, the abundance of halos in a universe with $\sigma=0.3$ is severely suppressed compared with that in the $\textit{Planck}$ cosmology. However, the story becomes different for the $n_\mathrm{s}$ dependence. As shown in the right panel, the amplitude of the PMF increases with the slope of the initial power spectrum for $\mu$ above $10^{-3}$. 

\revise{The results found here are consistent with analytical expectations in the EPS theory. In Appendix~\ref{app:EPS_prediction}, we evaluate the PMFs in EPS under varying cosmological parameters, and found that the PMFs are also nearly invariant under variations of $\Omega_{\rm M_0}$ and $\sigma_8$, but show a weak dependence on the $n_s$ parameter. Despite this, significant variations in the PMFs can only be observed with very large variations in the $n_s$ parameter (e.g., by a factor of two). For small variations (e.g., $\sim 10\%$) in the cosmology, it remains a very good approximation to describe the PMFs as universal.}

%\revise{For a theoretical perspective on the PMF’s universality under varying cosmological parameters, we provide a qualitative analysis using the Extended Press-Schechter (EPS) framework in Appendix~\ref{app:EPS_prediction}, which supports the robustness of the universality observed in our simulations.}

\section{The hierarchical origin of subhalos} \label{sec:hierarchical_origin}
The universal subhalo PMFs of different levels derived above provide a convenient tool to study the hierarchical origin of subhalos. In the following, we derive their level distribution, accretion redshift distribution, initial merger ratio distribution, and the merger rate at each level, starting from the model PMFs above. The derived properties are compared against direct measurements from the simulation when necessary.

\subsection{Accretion Level Distribution}
\begin{figure*}[htb!]
    \centering
    \includegraphics[width=0.48\textwidth]{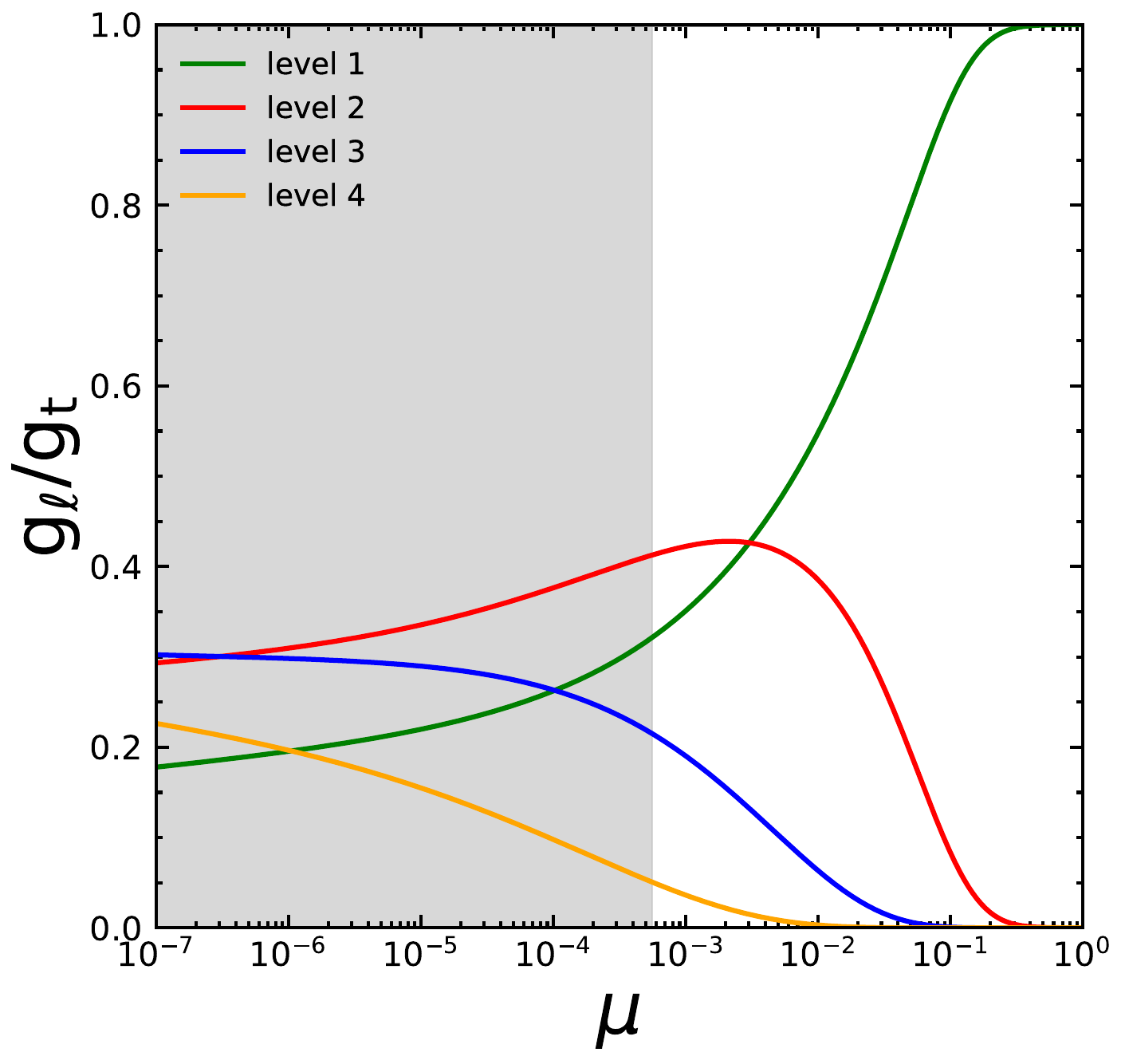}
    \includegraphics[width=0.48\textwidth]{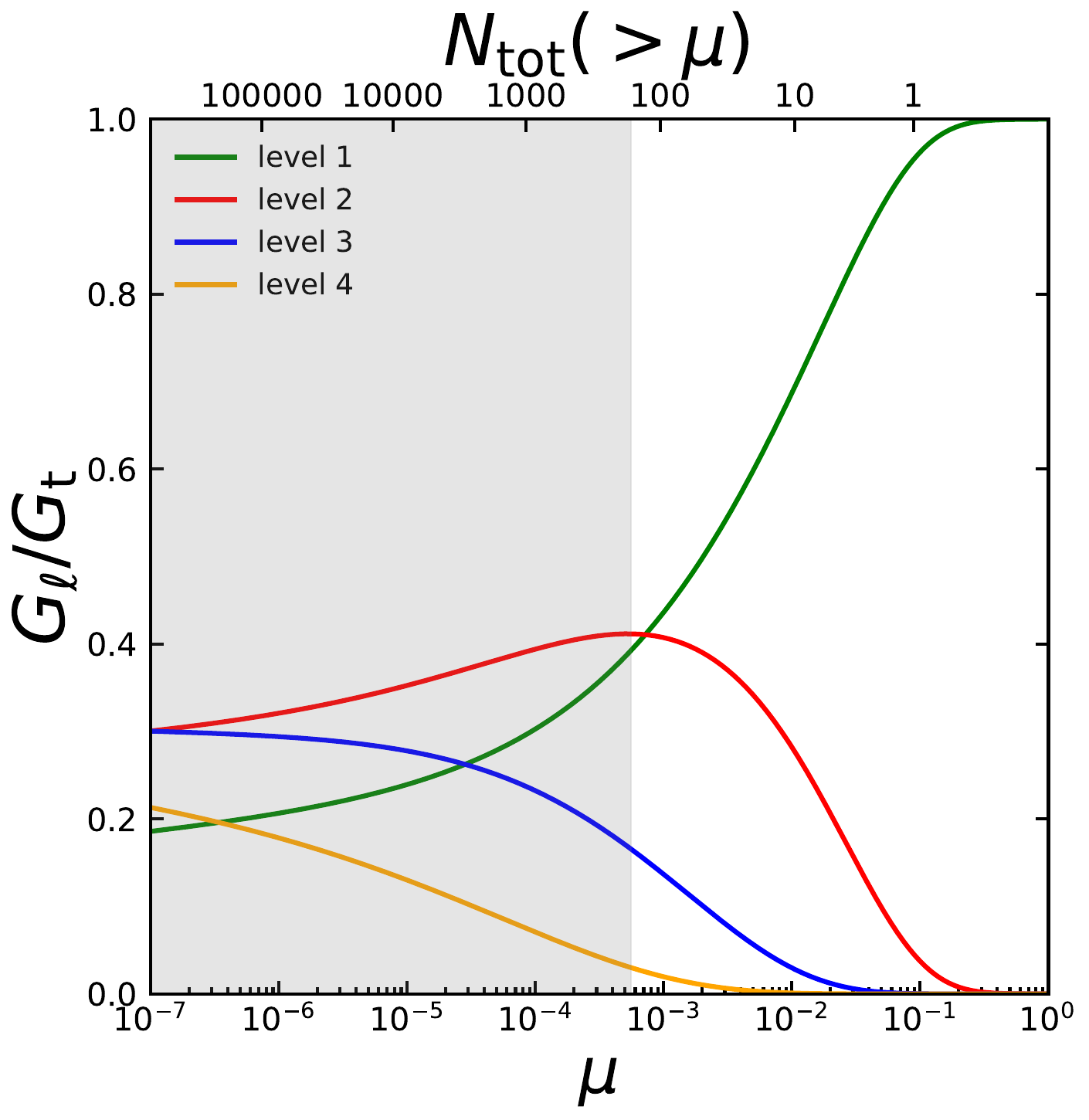}
    \caption{ The relative contributions from different levels to the total PMF. The left panel shows the contributions at a given $\mu$, while the right panel shows the cumulative contribution above a given $\mu$. The top x-axis of the right panel shows the number of subhalos in the total population with peak mass ratios exceeding a specified threshold $\mu$. The shaded gray region is not constrained by the simulation data and should be interpreted as extrapolations of the model.}
    \label{fig:hierarchy_compare}
\end{figure*}

For a subhalo of a given peak mass, the probability that it originates from a given accretion level is simply given by the rescaled PMFs. In the left panel of Figure~\ref{fig:hierarchy_compare}, we show the fractional contributions from subhalos of different levels to the total population as functions of the mass ratio.

Overall, subhalos with a high mass ratio are more likely to occupy lower levels. For the population of subhalos with $\mu$ greater than $10^{-2}$, at least half are classified as level-1 subhalos. As $\mu$ decreases, the level-1 subhalos become less dominant until the percentages of level-1 and level-2 subhalos in the total population equalize at $\mu=3\times10^{-3}$. Below this threshold, level-2 subhalos dominate and constitute approximately 40\% of the total population, until level-3 subhalos take over at $\mu<3\times10^{-7}$, constituting around 30 percent of the total population at the transition mass. For $\mu<10^{-6}$, less than 20 percent of subhalos are of level 1.

In the right panel of Fig.~\ref{fig:hierarchy_compare}, we show the cumulative PMFs, $G_\ell(\mu)\equiv N(>\mu)=\int_\mu^{\infty} g_\ell(\mu') \ud \ln \mu'$, normalized by the total cumulative PMF. This gives the fractional contribution from each level to the total progenitor counts above a given mass ratio $\mu$. Similar to the left panel, level-1 subhalos dominate the total population at the high $\mu$ end, while level-2 subhalos take over at $\mu<10^{-3}$. In the upper axis, we also show the cumulative number of subhalos, $G_t(\mu)$ at each $\mu$, corresponding to the average rank of a subhalo according to $\mu$. In terms of this peak mass rank, it can be seen that about 70 percent of the top 10 subhalos are from level 1. Among the top 100 subhalos, both level-1 and level-2 populations contribute about 40 percent. As more ranks are considered, level-2 subhalos start to dominate.

These results have important implications for studying the evolution of subhalos. In particular, low-mass subhalos are dominated by high-level ones which are subject to ``pre-processing'' within their previous hosts. These objects are also likely still orbiting inside their host subhalos rather than the central subhalo in the final host. To model their evolution, it is thus more important to study their orbits around, and the gravitational influence from their host subhalos. The evolution of high-level subhalos could also be more complex due to multibody interactions among the subhalo group in addition to their interactions with the host (sub)halo. For example, \citet{ludlow_unorthodox} found that some subhalos can be ejected far outside the host halo due to three-body interactions.

\subsection{Subhalo Accretion Rate at Each Level}
\label{sec:diff_merger}
In \citet{Dong22}, it was shown that the specific merger rate is the differential version of the PMF for level-1 subhalos. Here we provide an alternative derivation of this relation, applicable to subhalos of any level. Consider a subhalo of peak mass $m$ in a host halo of mass $M_h$, corresponding to $\mu_{\rm acc}=m/M_{h}$. At a later time, the host halo mass grows by $\delta M_h$, resulting in $\mu'_{\rm acc}=m/(M_h+\delta M_h)=\mu_{\rm acc}(1-\delta \ln M_h)$. The number of newly accreted subhalos with peak mass $[m, m+\ud m]$ is found as 
\begin{align}
    \delta N &=[g(\mu'_{\rm acc})-g(\mu_{\rm acc})]\ud \ln m\nonumber\\
    &=-\mu_{\rm acc} g'(\mu_{\rm acc})\delta\ln M_h \ud \ln m\nonumber\\
    &=-g'(\mu_{\rm acc})\delta \ln M_h \ud \mu,\nonumber
\end{align} where $g'(\mu_{\rm acc})=\ud g(\mu_{\rm acc})/\ud \mu_{\rm acc}$ is the derivative of the peak mass function. This immediately leads to the specific accretion rate as
\begin{align}
f(\mu_{\rm acc}) &\equiv \frac{\delta N}{\ud\mu_{\rm acc} \delta \ln M_h}\label{eq:mergerrate}\\
&=-g'(\mu_{\rm acc}).
\end{align}
The above equation can be applied to subhalos of any level, with 
\begin{equation}
    f_\ell(\mu_{\rm acc})=-g_\ell'(\mu_{\rm acc}).
\end{equation} It describes the accretion rate of level $\ell$ subhalos, originating from level $\ell-1$ subhalos, onto the host at a given redshift. %\revise{Note $\mu_{\rm acc}$ is the mass ratio relative to the host mass at the redshift of the accretion.}%
When $\ell=1$, the accretion rate becomes equivalent to the merger rate of progenitor halos.

\begin{figure*}[htb!]
    \centering
    \includegraphics[width=0.495\linewidth]{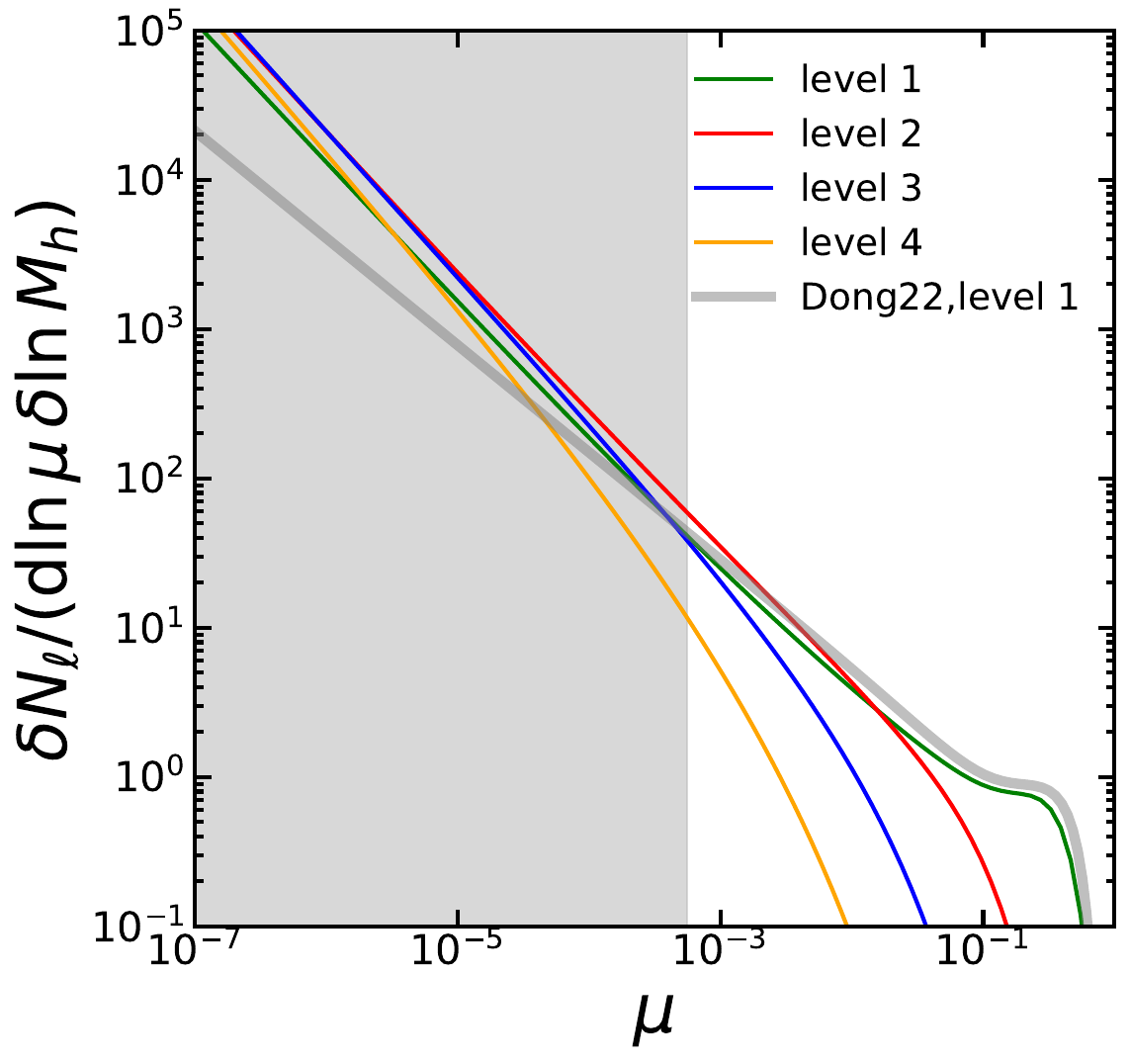}
    \includegraphics[width=0.48\linewidth]{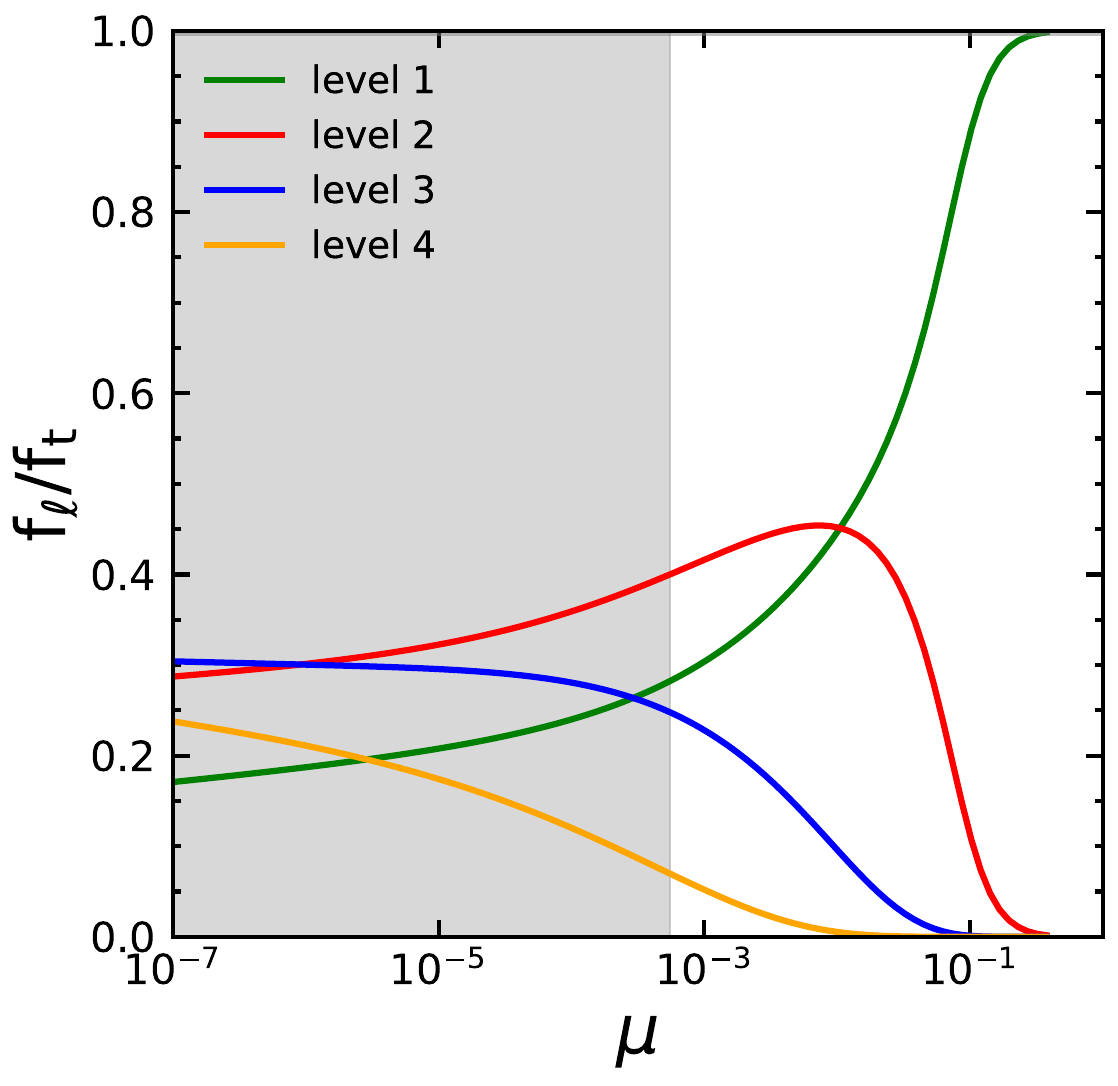}
    \caption{\textit{Left}: the specific accretion rate ($\mu_{\rm acc} f(\mu_{\rm acc})$ with $f$ defined in Equation~\eqref{eq:mergerrate}) for subhalos of different levels. The thin colored lines show predictions from our model. The thick gray line shows the model of \citet{Dong22} for level-1 subhalos, after correcting for the difference in the merger ratio definition (see footnote \ref{foot:ratio_correction}). \textit{Right:} The relative contribution from different levels to the total accretion rate. The gray shaded region in each plot highlight the mass range not constrained by our simulation data and should be interpreted as extrapolations of our model.}
    \label{fig:diff_case}
\end{figure*}
 
 The predicted specific accretion rate at each level and their contributions to the total merger rate are shown in the left and right panels of Figure~\ref{fig:diff_case}, respectively. In the left panel, the level-1 accretion rate is also compared against that of \citet{Dong22} for the halo merger rate.\footnote{The merger ratio in \citet{Dong22}, $\mu_{\rm Dong}$, is defined as the ratio between the progenitor masses prior to their merger, while in our case the ratio $\mu_{\rm acc}$ is between the subhalo peak mass and the host halo mass after the merger. We correct for the difference through
$\mu_{\rm acc}  = \frac{\mu_{\rm Dong}}{1+\mu_{\rm Dong}}$.\label{foot:ratio_correction}} Note both works are based on simulations that extend to $\mu_{\rm acc}\sim 5\times10^{-4}$. Above this mass ratio limit, the two models are largely consistent with each other. The slopes of the two models differ more significantly at the low mass end. This can be attributed to the different approaches in deriving the models. In \citet{Dong22}, the halo merger rate is measured directly from the halo merger tree under certain definitions for halo merger (e.g., in time and distance thresholds). Some post-processing of the tree is also required to account for fly-by and splash-back events. \revise{In our work, however, these events are automatically accounted for through the variation of the PMF at different time.} In addition, the virial masses prior to the merger time are adopted to define the merger ratio, in contrast to the peak mass and host bound mass used in our work. These operations could lead to systematic differences in the merger rate from that obtained in this work.

 In the right panel, the specific merger rate $f_\ell$ at each level has been normalized by the total rate, $f_t$, to show their relative contributions. The partitioning is similar to that shown in Figure~\ref{fig:hierarchy_compare}. The level-1 mergers still dominate the high $\mu$ region. With a decrease in $\mu$, contributions of higher-level mergers increase. At $\mu_{\rm acc}=0.017$, the specific merger rates for the first two levels, $f_1$ and $f_2$, equalize. It occurs at a higher ratio compared with the transition point between $g_1$ and $g_2$.
 
\subsection{Initial Merger Ratio Distribution}
\label{sec:initial_merger_ratio}
For high-level subhalos, their orbital and mass evolutions are initially determined by the host halo that they directly merge with, especially through their mass ratio relative to the initial host. The initial merger ratio distribution can be derived as a conditional probability distribution for a given final mass ratio, $\mu$, of level $\ell$ subhalos as 
\begin{align}
    p_{1|\ell}(\ln\xi|\mu)&\equiv \frac{\ud P(\xi|\mu)}{\ud\ln\xi}\\
    &=\frac{g_{\ell-1}(\mu/\xi)g_1(\beta \xi)}{g_{\ell}(\mu)}.
    \label{eq:initial_merger_ratio}
\end{align}
 Here, $\xi$ represents the \emph{peak} mass ratio between a level $\ell$ subhalo and its direct parent at level $\ell-1$. Note this ratio is a lower bound on the mass ratio between the subhalo and its direct host at the time of merger, as the direct host could still grow before it merges with the final halo.

\begin{figure*}[htb!]
    \centering
    \includegraphics[width=0.32\textwidth]{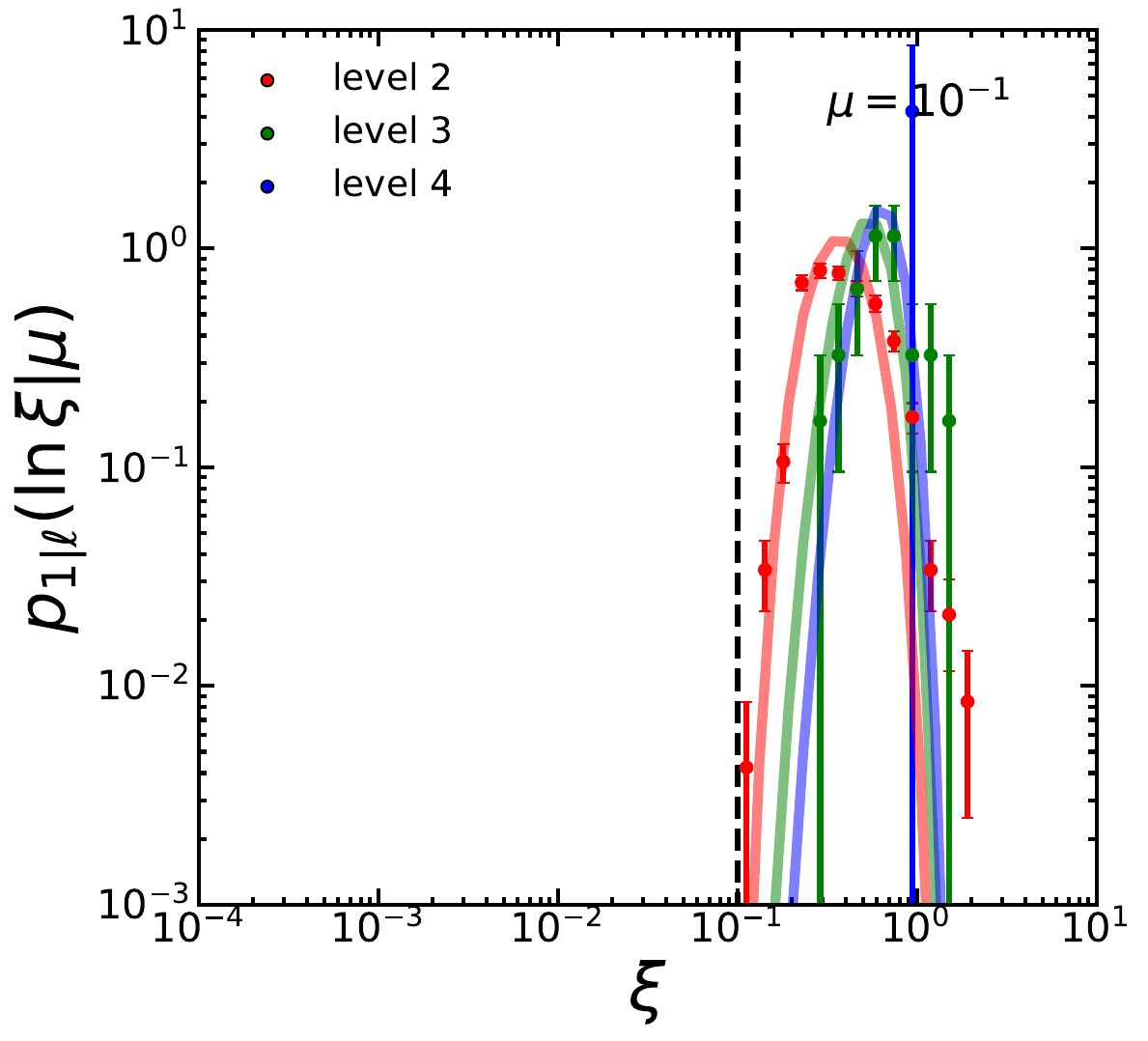}
    \includegraphics[width=0.32\textwidth]{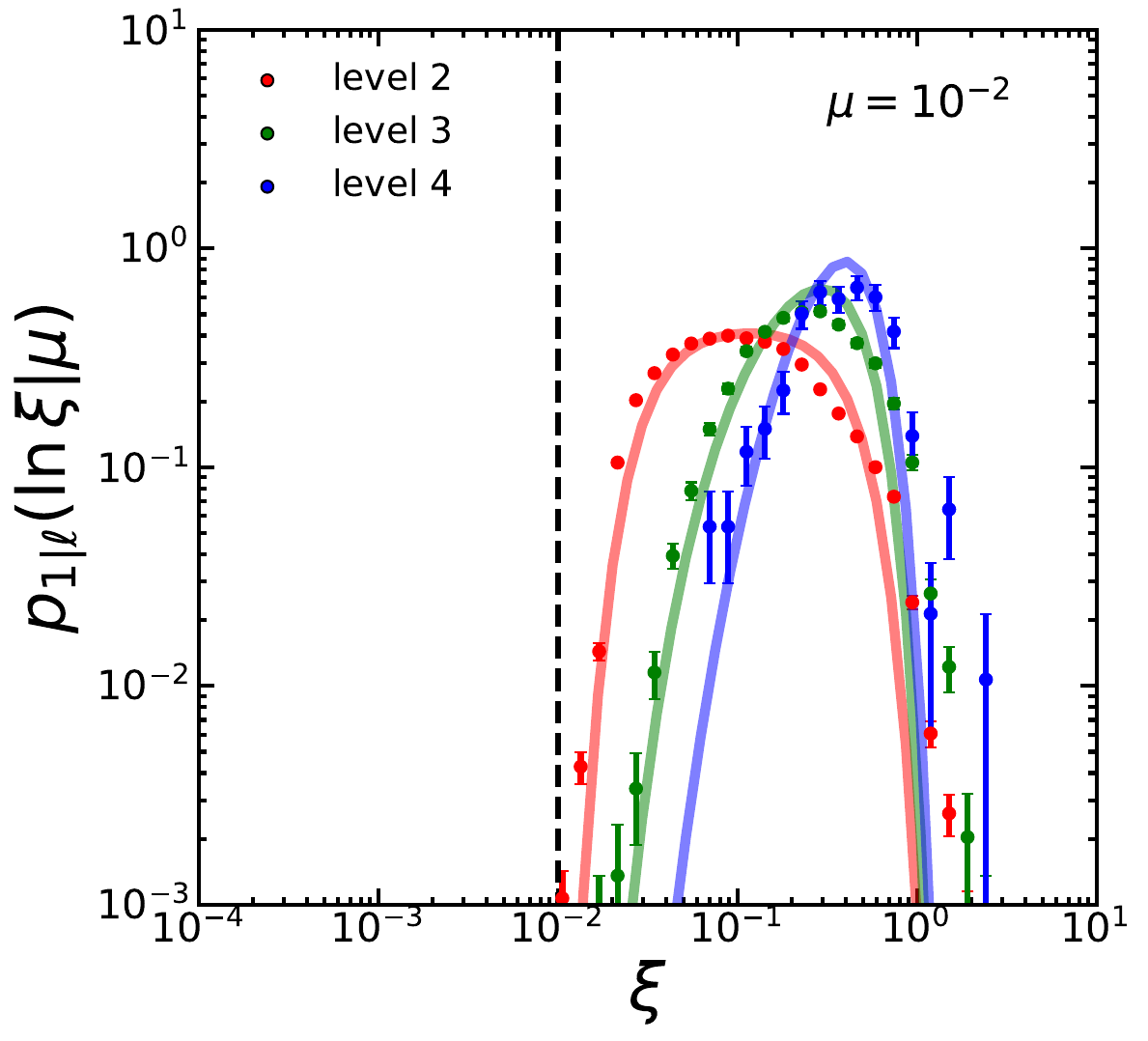}
    \includegraphics[width=0.32\textwidth]{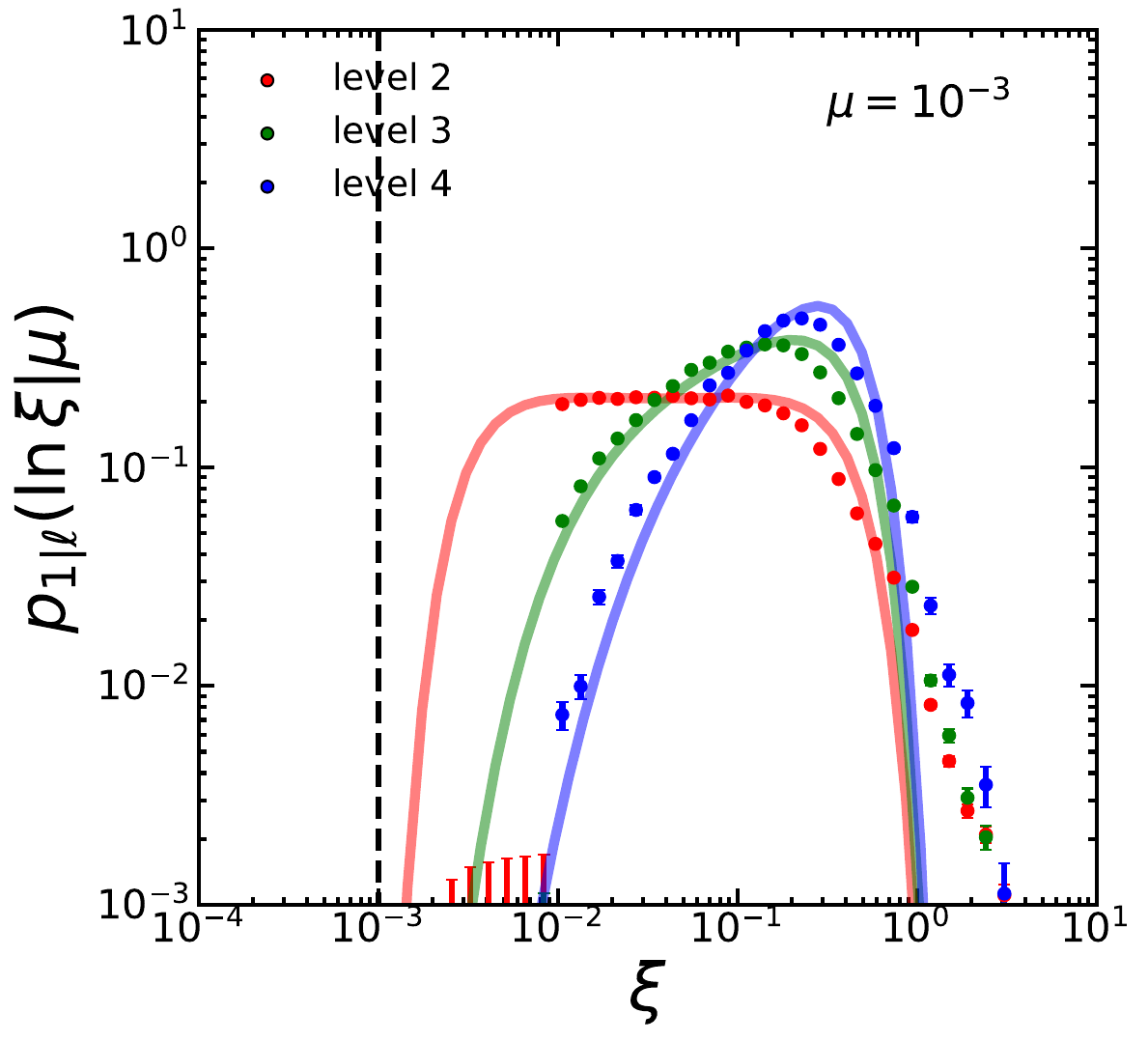}
     \caption{The probability distributions of the initial merger ratio, $\xi$, for subhalos of a given final mass ratio, $\mu$ (in different panels), and final subhalo level, $\ell$ (in different colors). The data points are measurements in the L600 simulation, while the solid lines show the model predictions. The vertical dashed curve highlights the $\mu$ value in each panel. \revise{The errorbars show the Poisson noise estimated from the total subhalo count in each initial merger ratio bin.}}
    \label{fig:initial_merger_ratio}
\end{figure*}

\begin{figure}[htb!]
    \centering
    \includegraphics[width=0.5\textwidth]{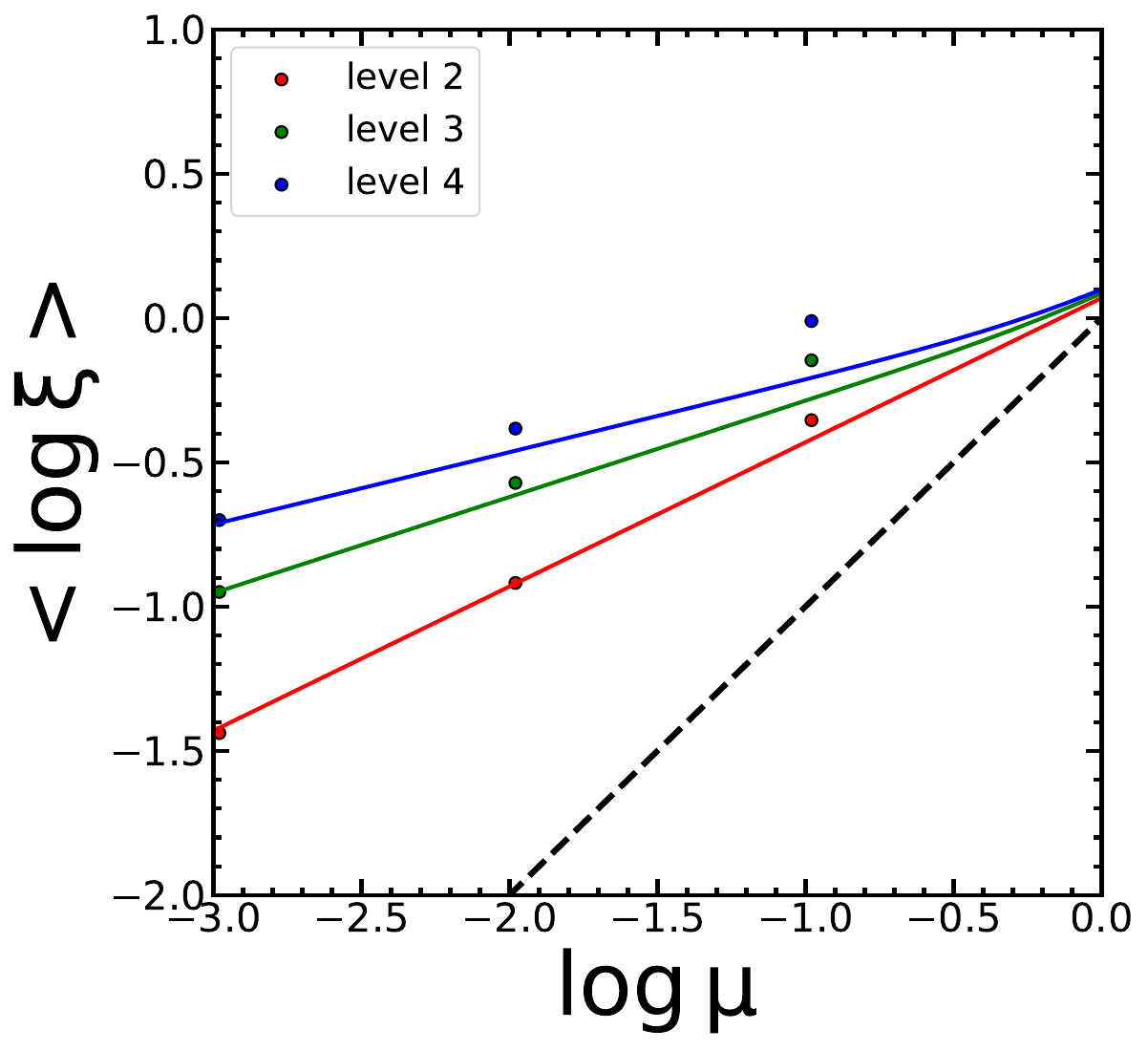}
     \caption{The predicted average of the logarithmic initial merger ratio compared with the logarithmic final merger ratio, for subhalos of different levels. The dashed line represents the 1:1 curve.}
    \label{fig:initial_merger_raito_mean}
\end{figure}

 Figure~\ref{fig:initial_merger_ratio} shows the predicted initial merger ratio distribution along with measurements from the L600 simulation. The predicted distribution closely matches the simulation data for the region $\xi<1$. From these three panels, we can observe two key features. First, for a given $\mu$, higher-level subhalos exhibit a more concentrated distribution of initial merger ratios, and their ratios tend to be shifted towards higher values. For example, the level-2 subhalos with $\mu=10^{-2}$ exhibit a relatively broad initial merger ratio distribution above $10^{-2}$. However, almost all of the level-4 subhalos with the same $\mu$ are accreted through major mergers with an initial merger ratio close to unity. Second, at a given $\ell$,  the initial merger ratio distribution is also more concentrated towards higher values when the final ratio, $\mu$, increases. These dependencies are also summarized in Figure~\ref{fig:initial_merger_raito_mean}, where we directly compare the average of the (logarithmic) initial merger ratio to the (logarithmic) final merger ratio. It can be seen that the average initial merger ratio is always higher than the final ratio and increases with the subhalo level and the final ratio. 
 
These results suggest that many higher-level subhalos may have experienced more violent evolution histories due to their higher initial merger ratios, even though they exhibit relatively small final peak mass ratios with respect to their host halos. 
 Our findings are consistent with the results of \citet{He24}, who found that many sub-subhalos are accreted through major mergers at high redshift in the simulation, resulting in very high mass loss rates responsible for their apparent disruptions. 
 
We note that some subhalos are observed with an excessively high ratio of $\xi>1$ in Figure~\ref{fig:initial_merger_ratio}, even though they constitute only a very small fraction (1\%–4\%) of the investigated population. The existence of these objects can be understood as due to systematics in building the merger tree. \hbt does not simply choose the most massive progenitor as the central subhalo during the merger, but relies on a kinematic distance to the \revise{center of mass of the host halo } to make the decision. Thus, a satellite may have a peak mass larger than that of the central. Furthermore, approximately 50\% of these high-merger-ratio subhalos have experienced central-satellite switching during their evolutionary history. During this switching process, they may acquire peak masses comparable to or even exceeding those of central subhalos. After all, the two progenitors are close in mass in a $\xi\sim 1$ merger, so it is always somewhat arbitrary to select one as central. We thus refrain from further discussing and modeling the distribution in this part.
 
\subsection{Accretion Redshift Distribution}
\label{sec:accretion_redshift_distribution}
For subhalos of a given final mass ratio $\mu_0$ in a host halo of mass $M_0$, its accretion rate at a previous redshift $z$ can be derived by rewriting Equation~\eqref{eq:mergerrate} as
\begin{equation}
    \frac{\delta N}{\ud \ln \mu_0 \delta z}=\mu_{\rm acc} f(\mu_{\rm acc}) \frac{\ud \ln M} {\ud z},
\end{equation} where $\mu_{\rm acc}=\mu_0/\tilde{M}(z)$ is the mass ratio at $z$, and $\tilde{M}(z)=M(z)/M_0$ is the normalized mass of the host halo at $z$. Normalizing the above accretion rate by the PMF of $\mu_0$ gives the accretion redshift distribution. For subhalos of level $\ell$, the probability density in redshift is
\begin{equation}
p_\ell(z|\mu_0,M_0)=\frac{\frac{\mu_0}{\tilde{M}(z)}f_\ell(\frac{\mu_0}{\tilde{M}(z)})}{g_\ell(\mu_0)} \frac{\mathrm{d}\ln \tilde{M}(z)}{\mathrm{d}z}.
    \label{eq:accrete_z_pdf}
\end{equation}

With universal models available in the literature to predict the mass accretion history, $\tilde{M}(z)$, Equation~\eqref{eq:accrete_z_pdf} can be evaluated explicitly. For example, \citet{Zhao09} developed a universal model for the median mass assembly history of dark matter halos with a given final mass at $z=z_0$~\citep[see also][]{vdB_MAH,Wechsler_MAH,Li_MAH,Ludlow_MAH,Liu24}. With this specific model, we can construct the accretion history of high-level subhalos analytically.

\begin{figure*}[htb!]
    \centering
    \includegraphics[width=0.95\textwidth]{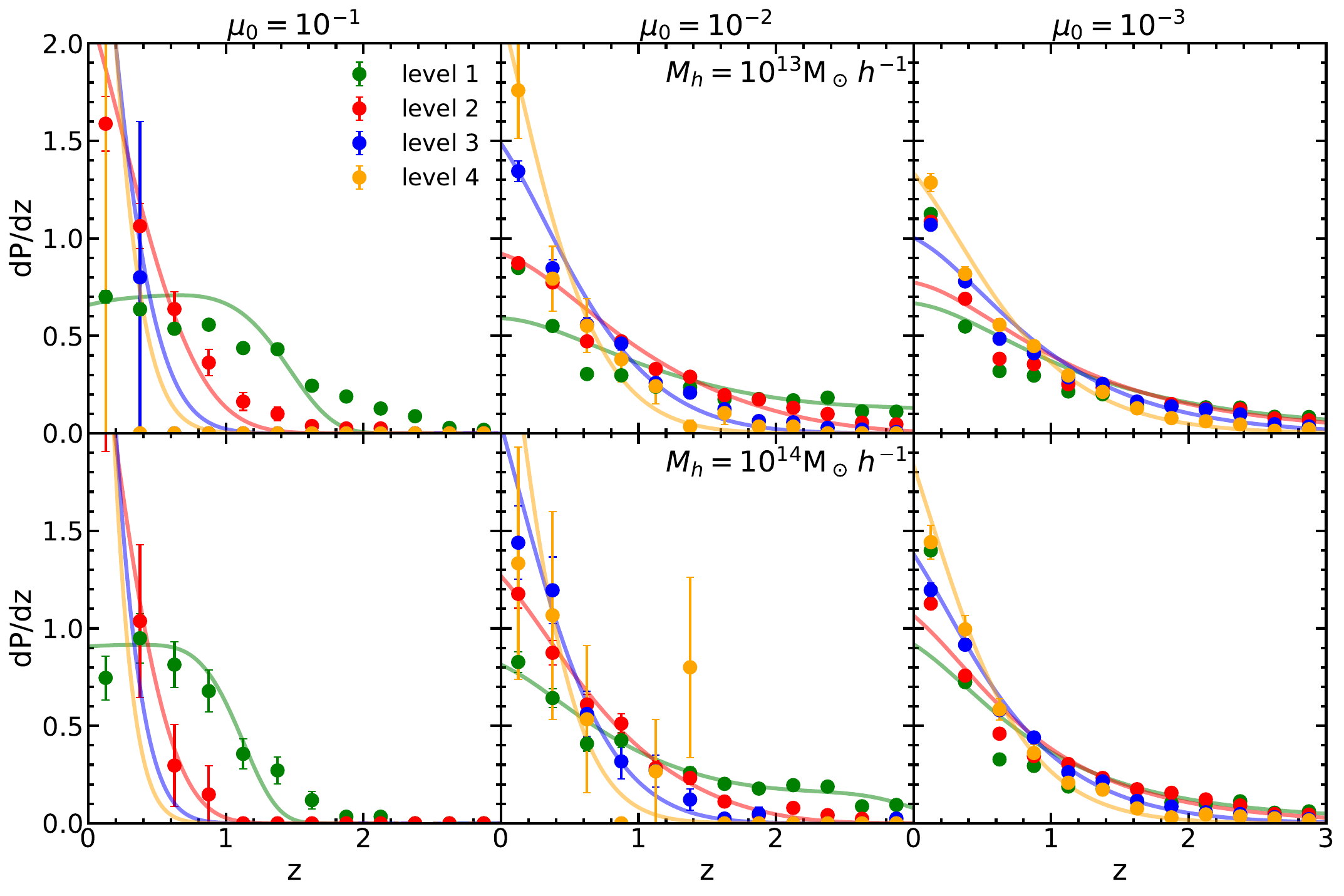}
    \caption{The accretion redshift distributions for subhalos of a given mass ratio, $\mu_0$, and host halo mass, $M_0$.  The data points show the simulation measurements in L600, while the solid lines show the model predictions. Different colors show the results for different subhalo levels. The top and bottom panels show the results for $M_0=10^{13}\msunh$ and $M_0=10^{14}\msunh$, respectively, while the different columns show the results for different $\mu_0$ as labeled. \revise{The errorbars show the Poisson noise estimated from the subhalo count in each redshift bin.}}
    \label{fig:acc_redshift}
\end{figure*}

Figure~\ref{fig:acc_redshift} compares the predicted redshift distributions against measurements from the L600 simulation. The predictions are in general agreement with the simulation results at the various $\ell$, $\mu_0$ and $M_0$ values explored. The results show that subhalos of different levels have different accretion redshift distributions, with higher-level subhalos tending to be accreted more recently. Subhalos with a higher $\mu_0$ also tend to be accreted more recently. These dependencies are intuitive to understand, as the progenitors of low redshift mergers have had more time to build up their subhalo levels and to grow in mass before the merger events. Comparing the distributions for different host halo masses, it can be seen that a higher mass host tends to accrete its subhalos more recently, consistent with the expectation that more massive halos build up later.

\section{Discussions: systematics in the PMF calculation}
\label{sec:discussion}
\subsection{Influences of Mass Definitions and Membership}
\label{sec:massdef&membership}
The computation of the PMF involves the determination of three key properties of a subhalo, including its peak mass, the assignment of it to a host halo, and the mass of the host halo. There are several popular choices on the definition of each of these properties. For the host halo mass, it is common to adopt the virial mass defined according to a certain density contrast within the corresponding virial radius. There are three widely used virial mass definitions, $M_{200c}$, $M_{200m}$ and $M_{\rm vir}$, corresponding to a virial density of 200 times the critical density, 200 times the mean density and that predicted by the spherical collapse model~\citep[e.g.,][]{Lacey93,ECF96, BryanNorman}, respectively. To be consistent with these spherical mass definitions, it is necessary to only include subhalos within the corresponding virial radius when computing the PMF. For the subhalo peak mass, the maximum virial mass along the evolutionary history according to one of the definitions can be used, besides the maximum bound mass. Alternatively, the virial mass at a certain infall time, e.g., the time when the subhalo first crosses the virial radius of the host halo, can be used as a peak mass.

We have tried various combinations of the different choices. Overall, the resulting mass functions can all be described by the double Schechter form. However, the level of universality varies depending on the mass and membership definitions. The optimal combination yielding the highest universality is the one adopted for our main results above, defining the peak mass as the maximum bound mass, the host halo mass as the sum of the subhalo bound masses, and the membership of subhalos according to the FoF catalog. We have also tried using the FoF mass as the host halo mass, which performs better than the virial masses while the universality is slightly worse than that using our fiducial definition.

\begin{figure*}[htb!]
    \centering
    \includegraphics[width=0.95\textwidth]{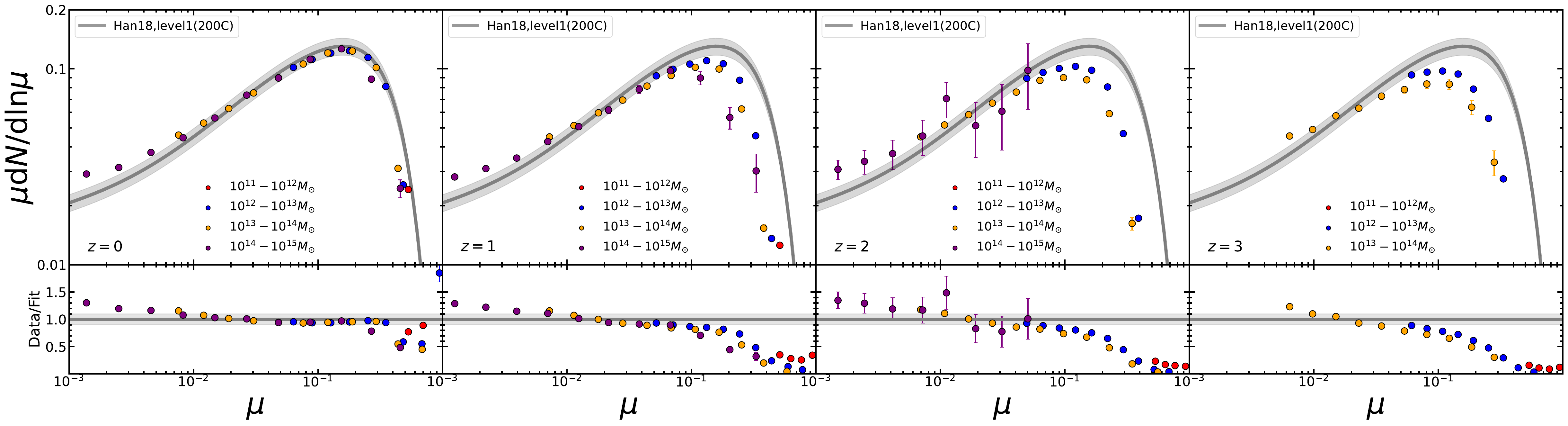}
    \includegraphics[width=0.95\textwidth]{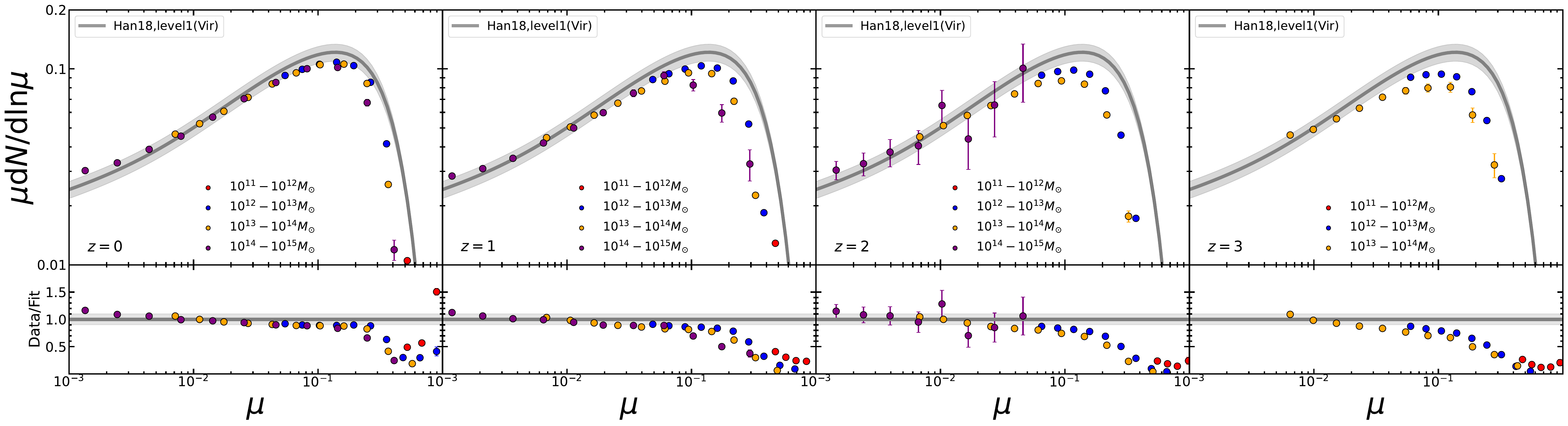}
    \caption{The PMFs obtained from L600 with host halo mass defined according to the virial definitions. Only subhalos (including disrupted ones) residing within the corresponding virial radius are counted. The top and bottom rows show results following the $M_{200c}$ and $M_{\rm vir}$ virial definitions, respectively. \revise{The errorbars show the uncertainty on the average PMF estimated from the Poisson noise in the total subhalo count in each bin.} Dots of different colors show results in different halo mass bins. The black solid line in each panel is the model $g_1$ converted from the best fitting $g_t$ in \citet{Han18} according to the recurrence relation.
    The gray-shaded region represents $\pm 10\%$ deviations from the model. The measurements from L600 have been corrected to the mass resolution of the Millennium-II simulation in order to compare against the \citet{Han18} model (see section~\ref{sec:unresolved_merger}).}.
    \label{fig:mem_comparison_R}
\end{figure*}

In Figure~\ref{fig:mem_comparison_R}, we show examples of two other combinations. In the top panels, the host halo mass is defined as $M_{200c}$, and subhalos are counted only if they are within the corresponding virial radius, $R_{200c}$, of the host halo. Disrupted subhalos are also counted with their locations traced by their most bound particles. The peak mass is defined as the maximum bound mass as in the fiducial combination. In the bottom panels, an alternative virial definition, $M_{\rm vir}$, is adopted. In both cases, a \revise{stronger } mass and redshift dependence at the high $\mu$ end can be observed compared with our fiducial combination. At a fixed host mass bin, the PMF tends to decrease with the increase in redshift at the high $\mu$ end, while the low $\mu$ distribution remains universal. This is equivalently observed as a decrease in the shoulder scale of the exponential tail of the PMF towards high redshifts. The evolution is stronger in more massive hosts and weaker in less massive hosts, leading to larger discrepancies among the PMFs at high redshifts.

\begin{figure*}[htb!]
    \centering
    \includegraphics[width=0.3\textwidth]{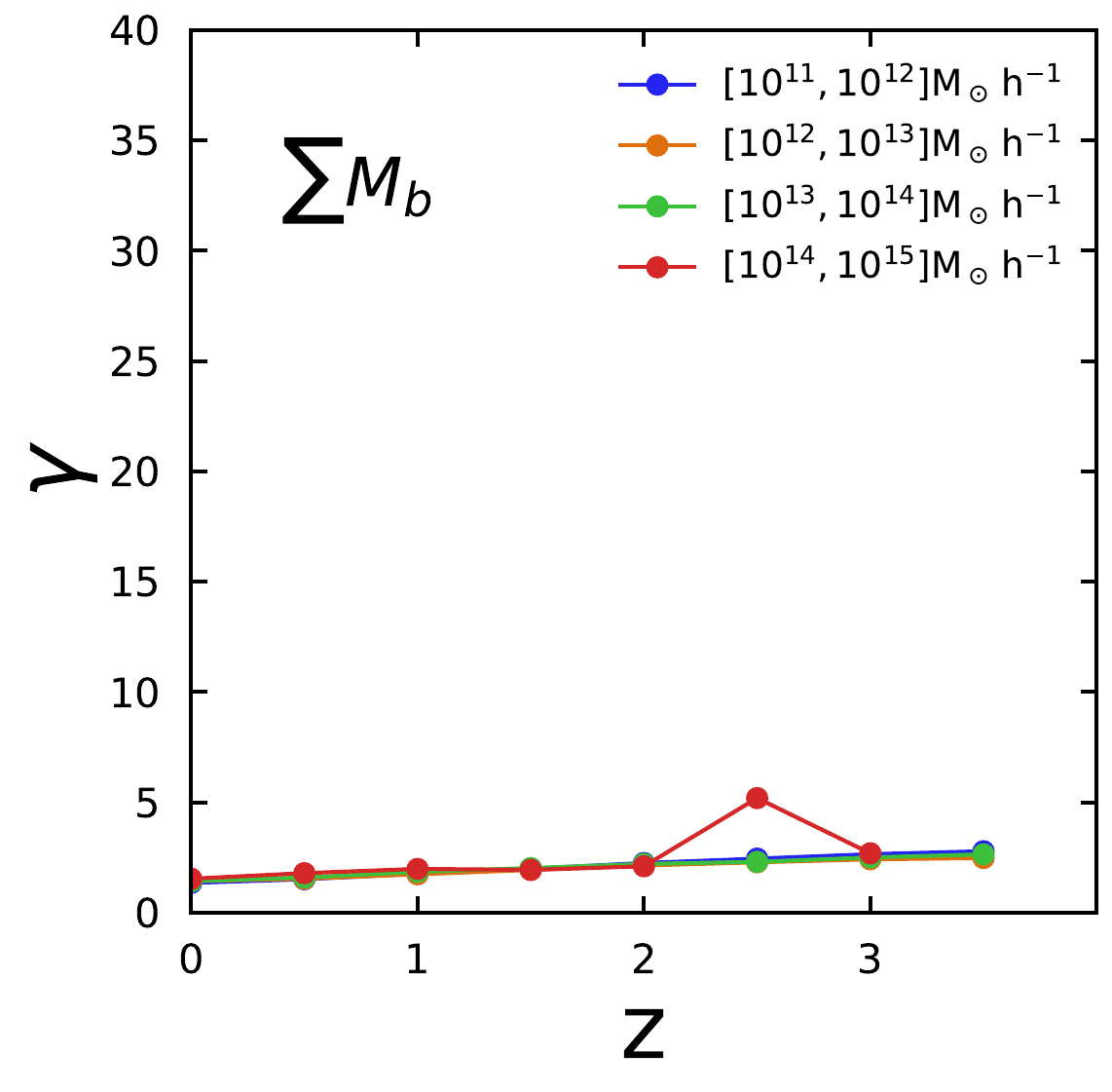}
    \includegraphics[width=0.3\textwidth]{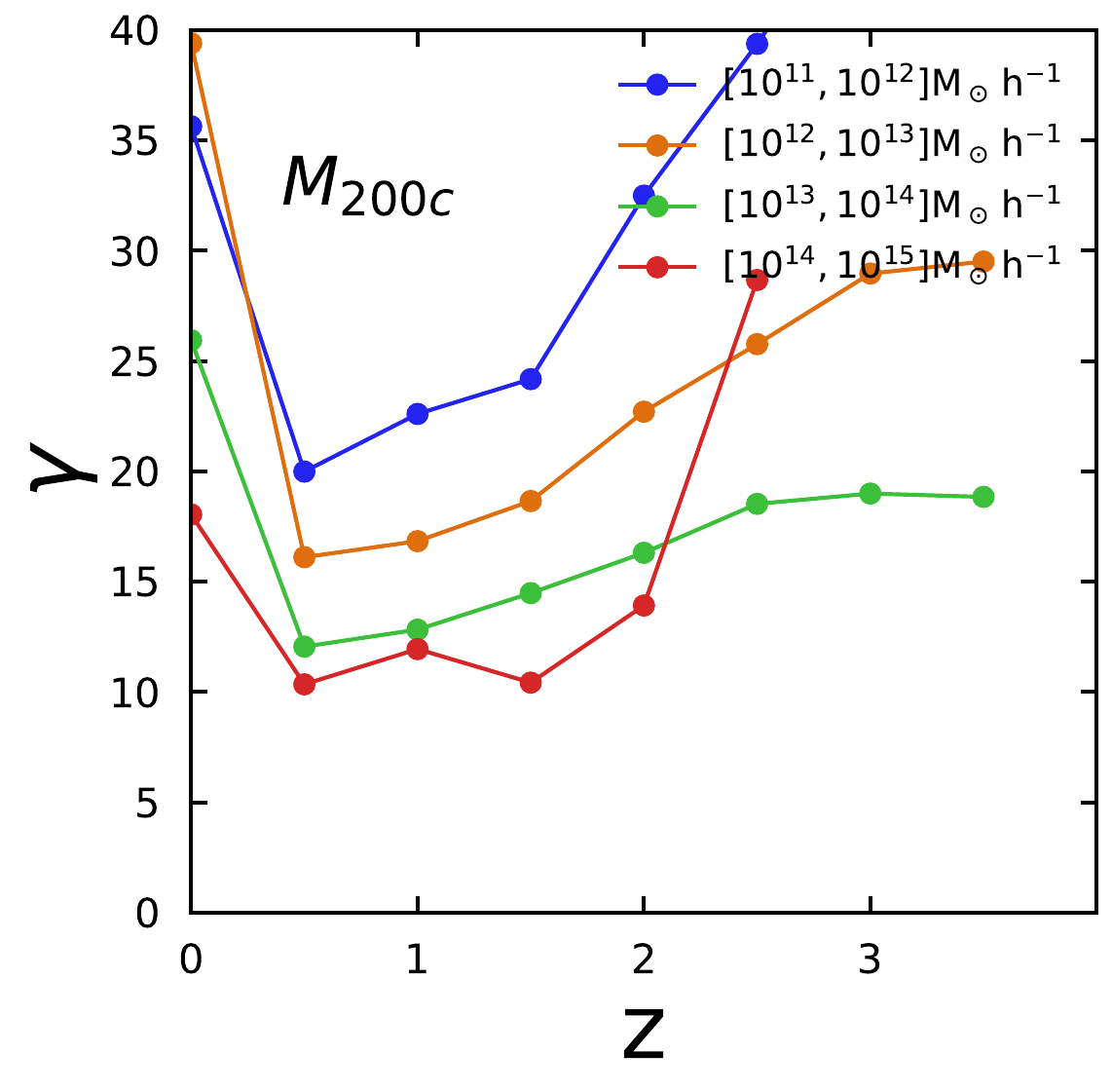}
    \includegraphics[width=0.3\textwidth]{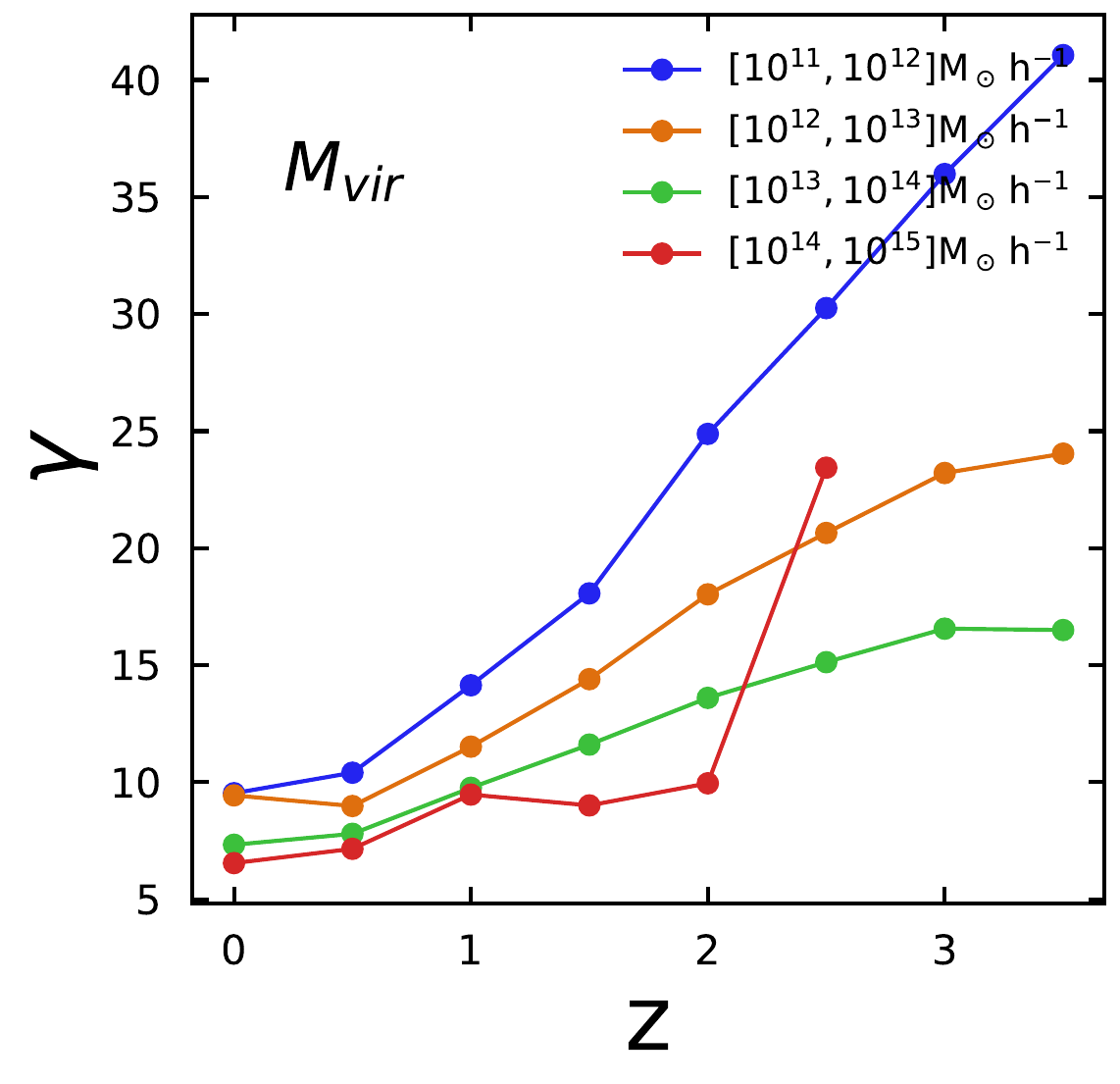}
    \caption{Dependence of the mass bias factor, $\gamma$, on host halo mass and redshift, measured in the L600 simulation. The left panel shows the result under the fiducial PMF definition. In the middle panel, the host halo mass is defined as $M_{\rm 200c}$ while subhalos are counted only if they are within $R_{\rm 200c}$. The right panel adopts $M_{\rm vir}$ and $R_{\rm vir}$ for the halo mass and boundary definition.}
    \label{fig:gamma_evo}
\end{figure*}

To get more insights into the degree of universality of the PMF under different definitions, we define a mass bias factor, $\gamma\equiv \sum \delta M_h/\delta\sum m_{\rm peak}$, as the ratio between the total increment in the halo mass and that in the sum of the subhalo peak mass for all the newly accreted subhalos.  To maintain good universality, it is necessary to require $\gamma$ to be independent of halo mass and redshift (see Equation~\eqref{eq:mergerrate}).The results, shown in Figure~\ref{fig:gamma_evo}, indicate that for our fiducial PMF definition, $\gamma$ remains nearly constant. The large fluctuation observed in $[10^{14},10^{15}]\msunh$ arises due to the lack of cluster-sized halos at high redshift (only one single halo is found in the mass bin). However, for the other two PMF definitions, both a strong halo mass dependence and a more pronounced redshift evolution are observed, along with systematically larger values. The membership based on the spherical overdensity excludes many newly accreted subhalos that have not yet crossed the radius of host halos but have already contributed a significant fraction of mass to the host. As a result, $\delta\sum m_{\rm peak}$ under such membership is systematically lower than that in the fiducial membership.

 These results are not difficult to understand. When two halos merge, the resulting FoF halo largely conserves the mass from both progenitors. Thus, when using the total bound mass of the FoF as a halo mass proxy, the increment in the halo mass can have a good scaling with the increment in progenitor mass. On the other hand, the spherical overdensity mass definitions do not \revise{reflect }  the morphology and extent of the merging halos, so that the scaling between the progenitor mass and halo mass is more complex. This complexity can also be understood as an inconsistency between the boundary of the host halo and the orbital extent of its member particles and subhalos. Recent halo boundary definitions that incorporate the splashback or growth envelope of a halo, e.g., the depletion radius~\citep{DepRad,Hongyu23}, could potentially provide more self-consistent boundaries for the study of halo mergers and the resulting mass and subhalo distribution~\citep{ZhouHan23,ZhouHan25,MWDep,Fong22}.

\subsection{Correcting Excessive Mass from Unresolved Subhalos}
\label{sec:unresolved_merger}
\begin{figure}[htb!]
    \centering
    \includegraphics[width=\linewidth]{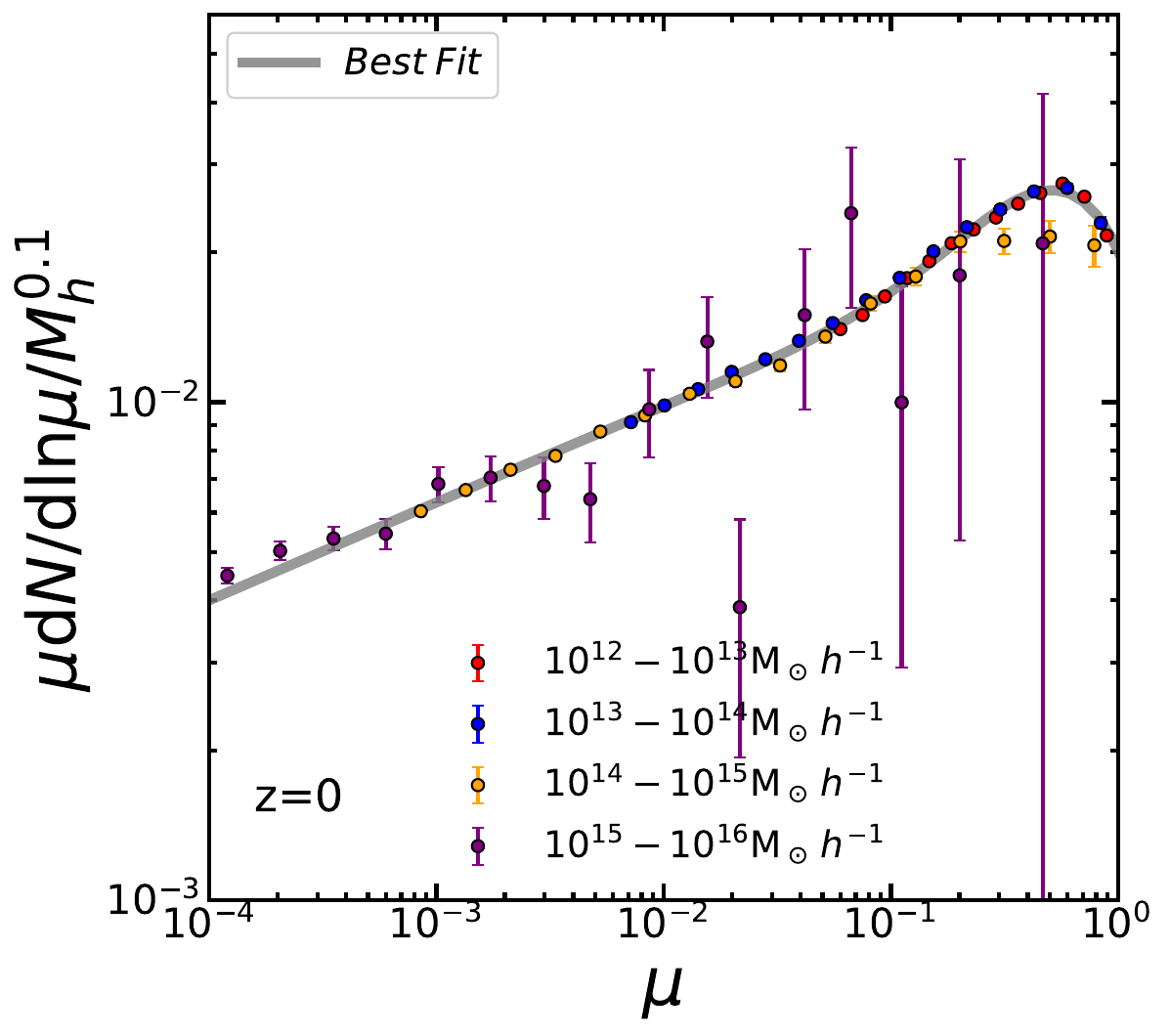}
    \caption{The final subhalo mass functions in L600 at $z=0$ after being rescaled by $(M_h/10^{10}\msunh)^{0.1}$. Dots of different colors show the results obatined from different halo mass bins. \revise{The errorbars show the uncertainty on the average final subhalo mass function estimated from the Poisson noise in the total subhalo count in each bin.} The gray solid line indicates the best fit (Equation~\eqref{eq:subMF_best_fit}).}
    \label{fig:subMF_Mb}
\end{figure}

\begin{figure*}[htb!]
    \centering
    \includegraphics[width=0.45\linewidth]{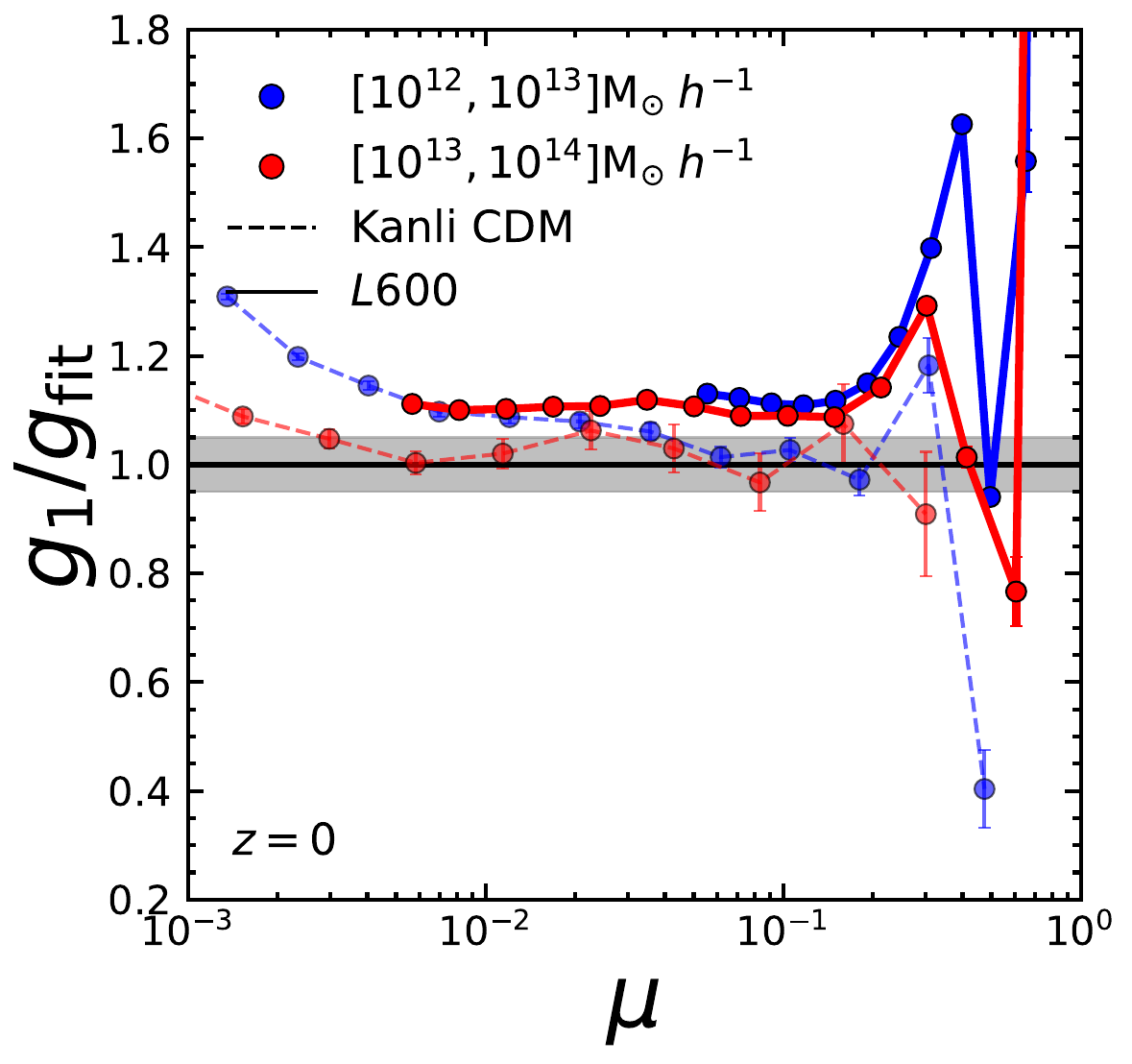}
    \includegraphics[width=0.45\linewidth]{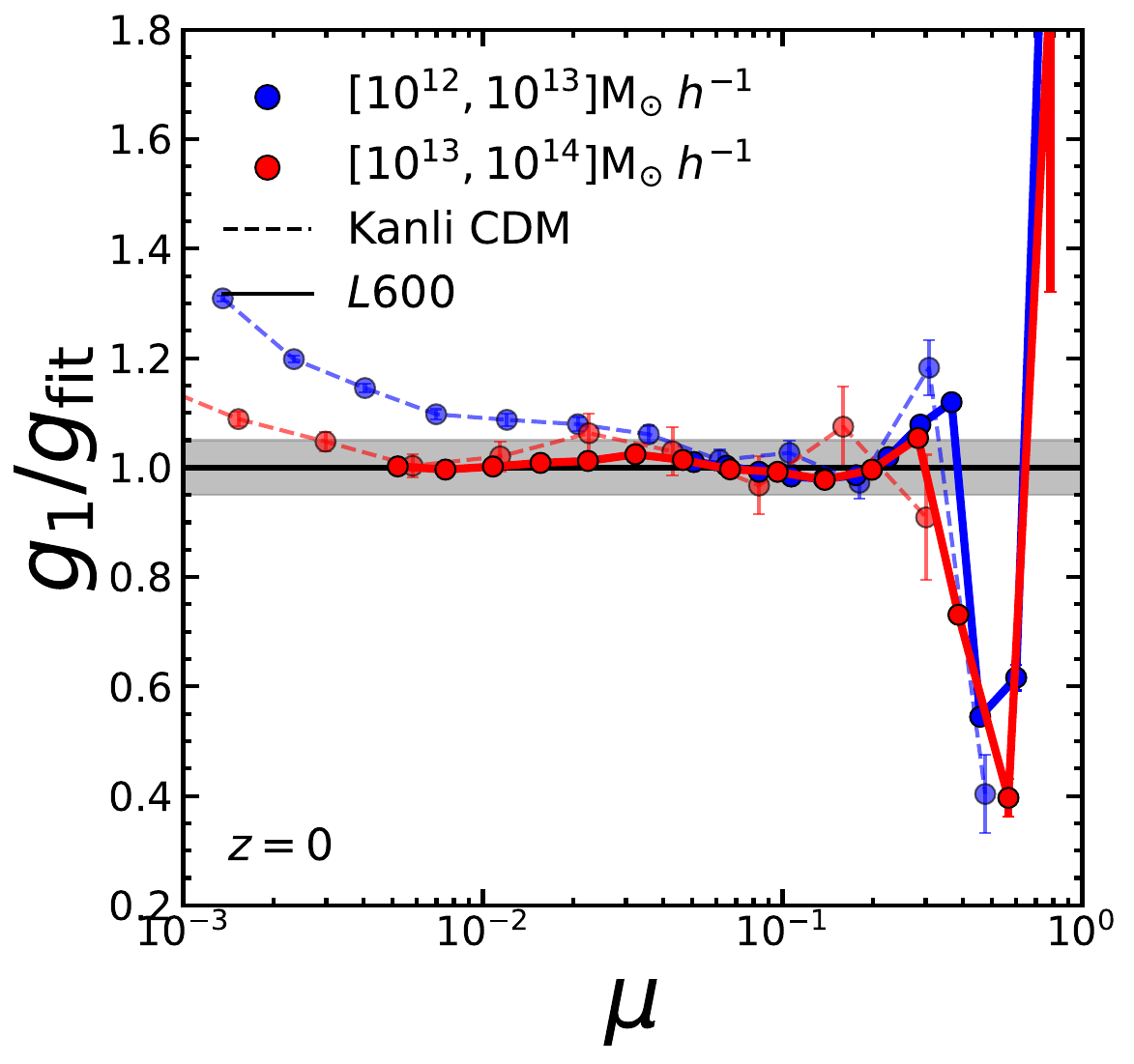}
    \caption{ \textit{Left}: Comparisons of $g_1$ obtained from the Kanli CDM simulation and L600 without peak mass correction. All of them have been normalized by the best-fitting formula in Equation~\eqref{eq:best_fit}. \revise{The thick solid and light dashed lines represent results from L600 and Kanli CDM, respectively. } In the plots, the PMFs are measured from two distinct host mass bins, $[10^{12}, 10^{13}]{\rm M_\odot}\,h^{-1}$ (blue points) and $[10^{13}, 10^{14}]{\rm M_\odot}\,h^{-1}$ (red points), in each simulation. Shaded regions indicate deviations of 5\% from the fitting formula. \textit{Right}: The comparisons of $g_1$ measured from two simulations after applying the mass correction to L600.}  
    \label{fig:mass_correct_comparisons}
\end{figure*}

In many subhalo finders including \hbt, the mass of a subhalo is defined to be the self-bound mass excluding particles belonging to other higher-level subhalos. This definition can suffer from a resolution dependence. For a simulation with a higher mass resolution, more embedded subhalos will be resolved and removed from the bound mass of their parents, resulting in a systematic reduction in the subhalo mass compared to a lower resolution run. More specifically, the subhalo mass, $m$, at a finite resolution can be related to the mass at infinite resolution, $m_0$, through
\begin{equation}
    m=m_0(1+f_{\rm unr}).
\end{equation}
Here $f_{\rm unr}=\int_0^{\mu_{\rm res}}\mu g_{\rm bound}(\mu)\ud \ln \mu$ is the mass fraction contained in unresolved subhalos, and $\mu_{\rm res}=m_{\rm res}/m_0\approx m_{\rm res}/m$ is the resolution limit in $\mu$, and $g_{\rm bound}$ is the \emph{final} subhalo mass function describing the bound mass distribution of subhalos. Note the mass ratio here is defined as the ratio between the bound mass of the subhalo and that of the host subhalo, while most previous works only compute the mass function relative to the virial mass of the host halo. In Fig.~\ref{fig:subMF_Mb}, we directly fit the final subhalo mass function measured from the L600 simulation, using the double Schechter function in Equation~\eqref{eq:db}. Following \citet{Han18}, the final subhalo mass functions from different halos are rescaled by $(M_h/10^{10}\msunh)^{0.1}$ before the fitting. The best-fitting parameters are as follows,
\begin{small}
    \begin{equation} (a_1^{\prime},a_2^{\prime},\alpha_1^{\prime},\alpha_2^{\prime},c^{\prime},d^{\prime})=(0.025,0.213,-0.80,0.54,2.47,0.96).
    \label{eq:subMF_best_fit}
    \end{equation}    
\end{small}
Considering the calculation of $f_{\rm unr}$ relies on the extrapolation of the final subhalo mass function down to zero mass, we can more conservatively correct the subhalo mass to a higher but finite resolution through  
\begin{align}
    m'&=\frac{m}{1+f_{\rm unr}}(1+f'_{\rm unr})
    \\&\approx m(1-\Delta f).
    \end{align}
Here $m'$ is the subhalo mass in a higher resolution run, $f'_{\rm unr}$ is the corresponding unresolved fraction, and $\Delta=f_{\rm unr}-f'_{\rm unr}$ is the mass fraction resolved in between the two resolutions.  As shown in Fig.~\ref{fig:mass_correct_comparisons}, $g_1$ measured from the L600 simulation ($5.5\times10^{8}{\rm M_\odot}\;h^{-1}$) shows at least a 10\% deviation from that obtained in the Kanli CDM simulation ($1.01\times10^{7}{\rm M_\odot}\;h^{-1}$). On the other hand, correcting the subhalo masses in L600 to higher resolution nearly eliminates the deviation.

\section{Conclusion}
\label{sec:conclusion}
In this work, we have carried out a detailed analysis on the hierarchical origin of dark matter subhalos, using a set of N-body simulations. Applying the \hbt subhalo finder to the simulations, we classify subhalos into different levels according to their merger histories, and use the peak mass function (PMF) of each level to summarize their hierarchical origins, in which the peak mass of a subhalo is used as a proxy for the progenitor mass. Our main findings are summarized below.

\begin{itemize}
    \item The \revise{PMF } is very close to a universal function across halo mass and redshift. A single double-Schechter fit can describe the level-1 subhalo PMF to $10\%$ accuracy down to a peak-to-host mass ratio of $\mu=5\times10^{-4}$, for the halo mass range of $10^{11}\msunh-10^{15}\msunh$ and redshift range of $z=0-3$ covered in our analysis.
    \item The PMF is insensitive to variations in the cosmological parameters including $\Omega_m$, $\sigma_8$ and $n_s$ around the concordance $\Lambda$CDM model. Small variations in these parameters by (e.g., at 10\% level) produces negligible differences in the PMF according to both simulations and EPS predictions. Among them, the $n_s$ parameter shows the highest sensitivity in shaping the PMF.
    \item The PMFs for high-level subhalos can be derived from that of the level-1 PMF through self-convolutions. The analytical model requires only one extra parameter besides those specifying the level-1 PMF, and provides a good match to the simulation results across mass and redshift. A small deviation is observed in the highest-level PMF covered in our analysis, indicating a weak evolution of the PMF over mass and redshift.
    \item The level of universality of the PMF differs for different definitions of the host halo mass, subhalo mass and subhalo membership. We find that defining the subhalo mass with the peak bound mass, the halo mass with the total bound mass in the FoF, and counting subhalos according to the FoF membership results in the optimal universality. Defining the halo mass using the virial definitions exhibits stronger evolutions over halo mass and redshift at the high mass end. To achieve better universality, we have also developed a method to correct for the resolution dependence in the subhalo mass. 
\end{itemize}

The analytical PMF model developed in this work provides a convenient tool for studying the progenitor populations of dark matter halos, as well as for populating halo progenitors in Monte-Carlo models. It is much simpler than the classical EPS theory, while at the same time provides a well-calibrated description of the simulation data which may not be achievable in simple EPS models. The model itself is a concise summary of the simulation results, and thus could serve as a starting point for further theoretical understanding of the hierarchical structure formation process. The derivation of the high-level PMFs through self-convolution of the level-1 PMF is also a direct reflection of the at least approximate \emph{self-similarity} of the halo merger tree, such that a truncation of a tree at any branch results in a smaller tree of the same analytical properties.

Starting from the universal PMF model, we further derive a series of analytical properties on the hierarchical origins of subhalos. We find
\begin{itemize}
    \item At the high mass end, the progenitor population are dominated by level-1 progenitors, while higher-level progenitors dominate at progressively lower masses. Among the top 100 most massive progenitors (or $\mu>10^{-3}$), level-1 and level-2 progenitors both contribute $\approx 40\%$. 
    \item At a fixed mass ratio at accretion time, the subhalo accretion rate at each level is proportional to the mass accretion rate of the host halo. This result generalizes the findings of \citet{Dong22} on the universal specific merger rate of dark matter halos to high-level subhalos. In line with PMF decomposition, the accretion rate of level-1 subhalos dominates at the high mass end ($\mu>10^{-2}$), while those of higher-level subhalos dominate at progressively lower masses.
    \item The peak mass ratio of a high-level subhalo relative to its initial host can be substantially larger than the ratio relative to its final host. The average initial merger ratio increases with both the final ratio and the subhalo level. For example, at a final ratio of $\mu=10^{-3}$, almost all subhalos of level-4 have an initial ratio above $10^{-2}$, with a most-probable ratio of $\sim0.3$. This means high-level subhalos are more likely to originate from major mergers when they first become subhalos.
    \item The accretion redshift distribution depends on the subhalo level, peak mass and host mass. Higher-level subhalos and those in more massive host halos tend to be accreted more recently, when the progenitors have had enough time to build up their own subhalo population. On the other hand, subhalos of a lower peak mass ratio tend to be accreted from a higher redshift.   
\end{itemize}
Because the evolution of a subhalo is determined by the properties of its direct host at least in the beginning, the above results highlight the necessity for correctly identifying the true initial condition of the subhalo merger, instead of relating their evolutions solely to the properties of the final host. In upcoming works, we will study the subsequent evolution and fate of the hierarchical subhalo population of different levels in detail.

\begin{acknowledgments}
This work is supported by National Key R\&D Program of China (2023YFA1607800, 2023YFA1607801), NSFC(12303003), 111 project (No.\ B20019), and the science research grants from the China Manned Space Project (No.\ CMS-CSST-2021-A03). We acknowledge the sponsorship from the Yangyang Development Fund. The computation of this work is done on the \textsc{Gravity} supercomputer at the Department of Astronomy, Shanghai Jiao Tong University. 
\end{acknowledgments}

\bibliography{example}{}
\bibliographystyle{aasjournal}

\appendix
\revise{\section{Halo-to-halo variation in the PMF}
\label{app:halo-to-halo_variation}
\begin{figure}
    \centering
    \includegraphics[width=0.465\linewidth]{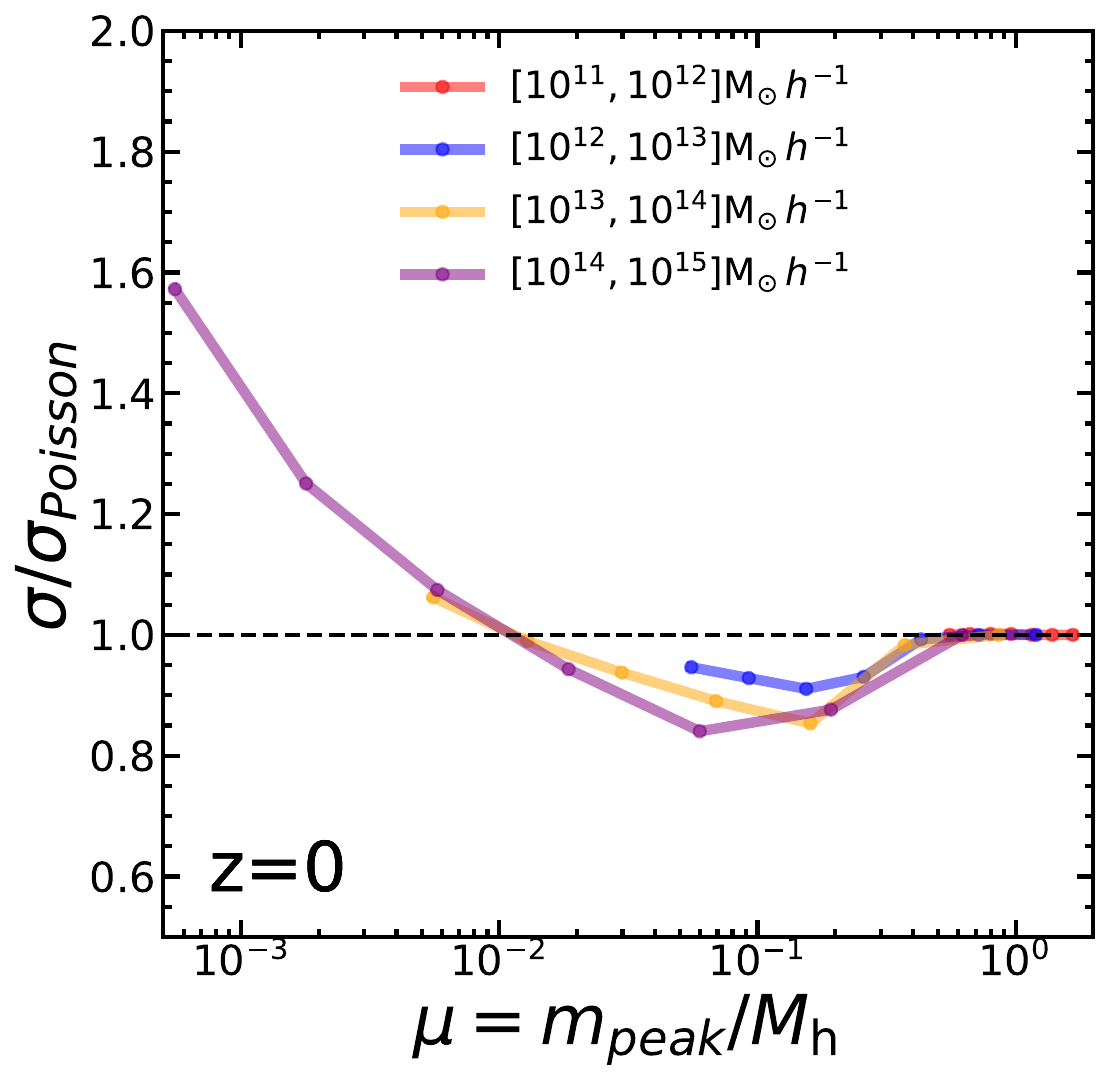}
    \caption{ The ratio between the standard deviation of the PMFs among different halos and the Poission noise in the PMF of an average halo. Different colors show the results in different halo mass bins as labeled. The dashed line represents $\sigma/\sigma_{\rm Poisson}=1$.}
    \label{fig:non-poissonity} 
\end{figure}

To investigate the significance of halo-to-halo variation in the PMFs, we compare %Halo-to-halo variation can be an important factor to further investigate the universality of the PMFs. In this appendix, we show the non-Poissonity of $g_1$ among different halos. We investigate it by showing the ratio between 
the standard deviation of subhalo counts ($\sigma=\sqrt{\sum_{i=1}^{N_{\rm h}} \frac{(N_{\rm sub,i}-\bar{N}_{\rm sub})^2}{N_{\rm h}}}$) among different halos and the Poisson noise ($\sigma_{\rm Poisson}=\sqrt{\bar{N}_{sub}}$).

In Figure~\ref{fig:non-poissonity}, we plot the ratio of the measured scatter in the level-1 PMFs to the Poisson expectation, $\sigma/\sigma_{\mathrm{Poisson}}$, as a function of mass ratio $\mu$. For low mass ratios ($\mu \lesssim 10^{-2}$), the scatter is super-Poissonian ($\sigma/\sigma_{\mathrm{Poisson}} > 1$), indicating a greater variability than that predicted by a Poisson distribution. In the intermediate range ($10^{-2} \leq \mu \leq 2 \times 10^{-1}$), the scatter is sub-Poissonian ($\sigma/\sigma_{\mathrm{Poisson}} < 1$). For higher mass ratios ($\mu > 0.2$), the scatter aligns closely with the Poisson expectation ($\sigma/\sigma_{\mathrm{Poisson}} \approx 1$). These halo-to-halo variations are consistent with previous studies \citep[e.g.][]{Bk10, Jiang&vdb_3, Chua}. Additionally, the PMF scatter exhibits a mild dependence on halo mass, with the sub-Poissonian behavior more pronounced in massive halos. Finally, across the range of peak mass ratio $\mu$ considered in the paper, the observed scatter does not exceed 1.6 times the Poisson error.}

\section{The cosmology dependence of the PMF in the EPS theory}
\label{app:EPS_prediction}
\begin{figure*}[htb!]
    \centering
    \includegraphics[width=0.49\linewidth]{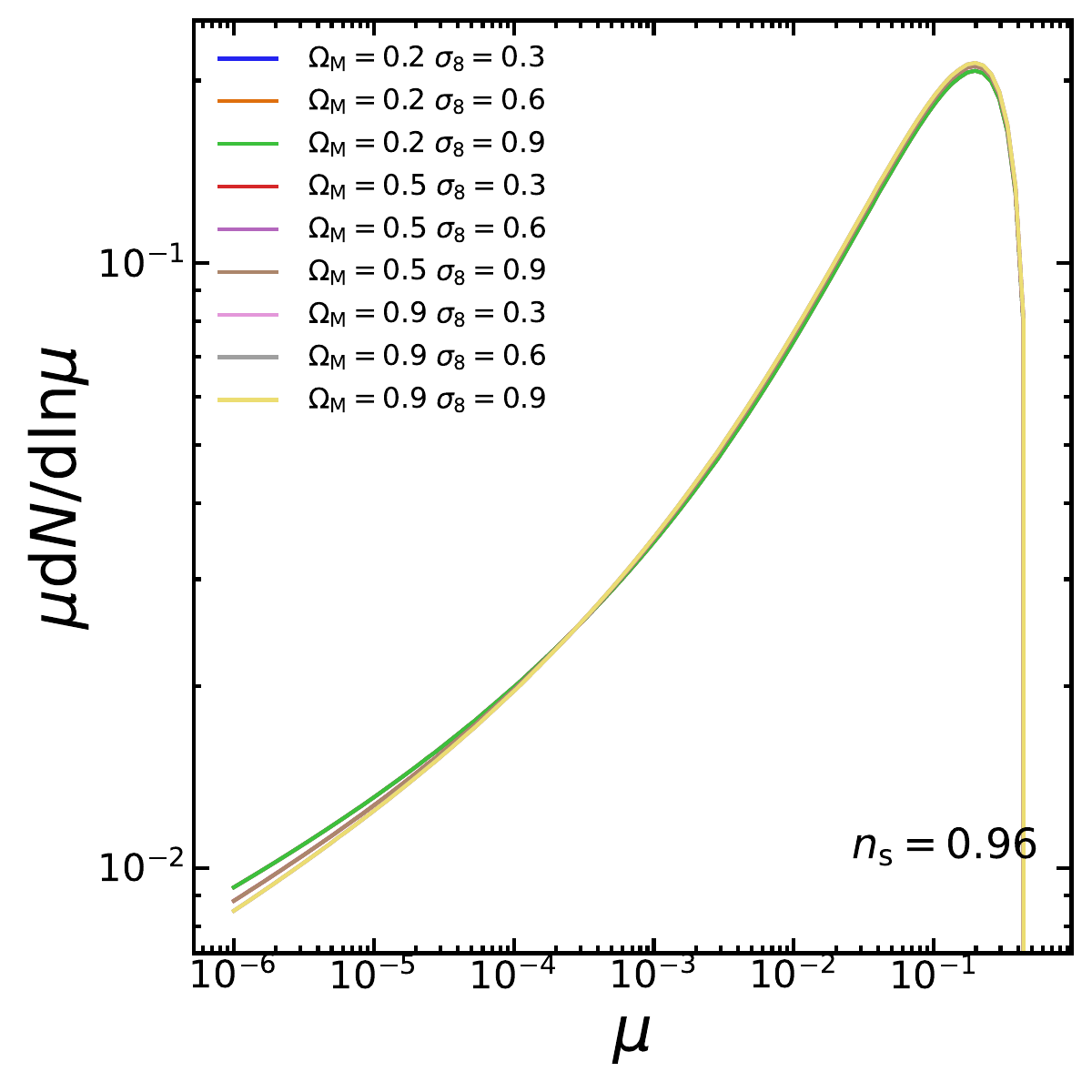}
    \includegraphics[width=0.49\linewidth]{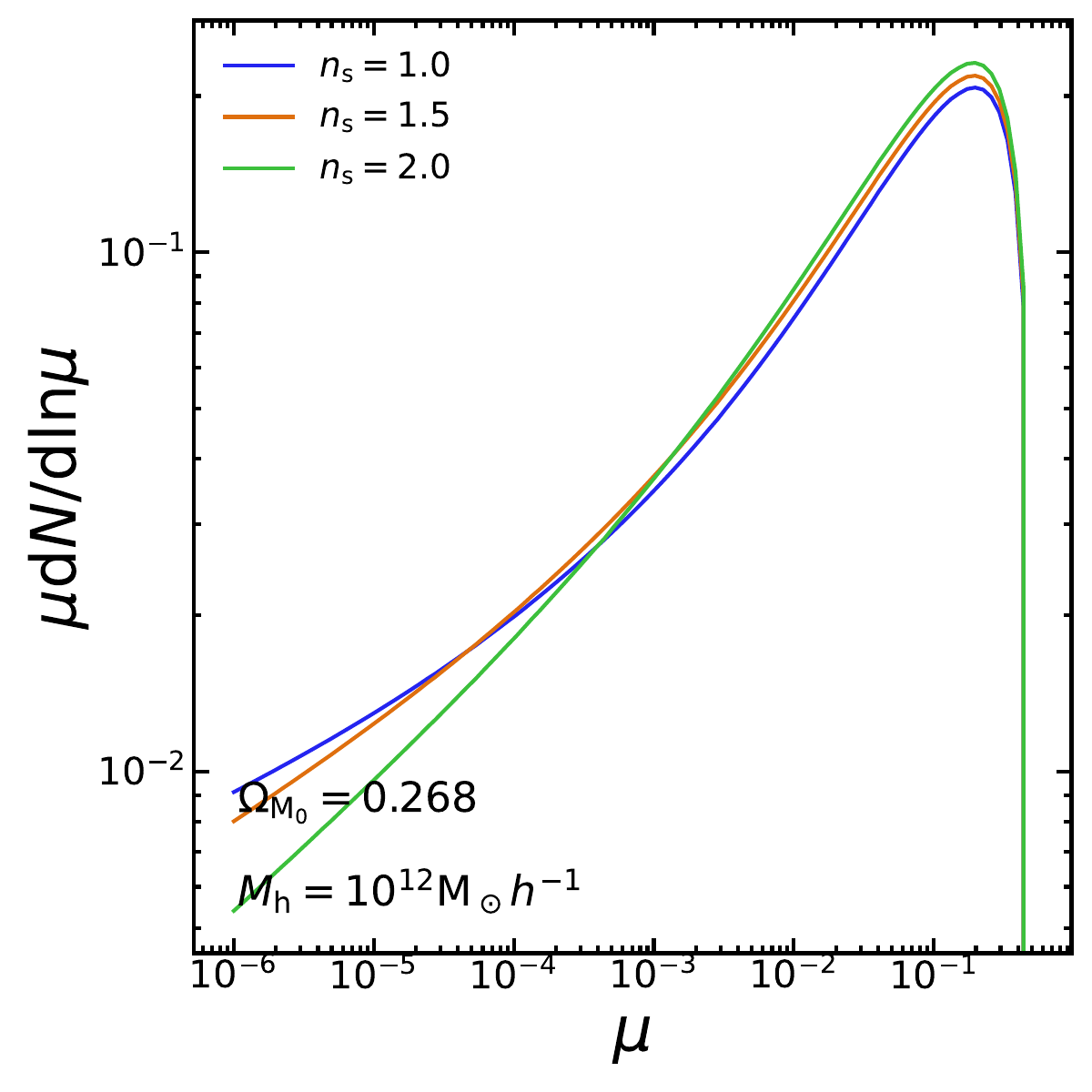}
    \caption{The level-1 PMF predicted from the EPS formalism in host halos of mass $M_h=10^{12}\msunh$, for different cosmological parameters. The left panel shows the results when varying $\sigma_8$ and $\Omega_{\rm M_0}$ away from the concordance $\Lambda$CDM cosmology. The right panel shows the result under different $n_{\rm s}$ with $\Omega_{\rm M_0}=0.268$ and $\sigma_8=0.863$.}
    \label{fig:EPS_PMF}
\end{figure*}

 \revise{In this appendix, we provide a theoretical interpretation for the PMF dependence on different cosmological parameters with the extended Press–Schechter (EPS) formalism . } We follow \citet{Dong22} to calculate the specific merger rate using the EPS theory (their Equation (11)) and then convert the result to the PMF (their Equation (4)). We use the \citet{Eisen_Hu98} transfer function to calculate the linear power spectrum and adopt a top-hat filter in Fourier space to compute the variance of fluctuation on a given scale. The predictions for the PMFs under different cosmological parameters are presented in Figure \ref{fig:EPS_PMF}. 

In the left panel, the predicted PMFs are nearly universal for $\mu$ values greater than $10^{-3}$, regardless of variations in $\Omega_{M_0}$ and $\sigma_8$. The slopes of the predicted PMFs show slight differences below $10^{-3}$, where larger $\Omega_{M_0}$ values result in steeper slopes. This cosmology dependence also suggests a weak redshift dependence in a given cosmology due to the evolution of the matter density and fluctuation amplitude with redshift. Verifying this residual dependence requires higher-resolution simulations than those used in the current work.

The right panel of Figure~\ref{fig:EPS_PMF} shows the predicted PMFs for different $n_{\rm s}$. Again, the ranges above and below $\mu\sim 10^{-3}$ exhibit different properties. Generally, a larger $n_{\rm s}$  results in a steeper slope at the very low mass end. This can be understood as the $n_s$ parameter controls the relative abundance of fluctuations on different scales and hence determines the shape of the halo mass function, which in turn affects the shape of the progenitor (peak) mass function. This finding is also consistent with the results reported by \citet{Yang11}, where the slope of the subhalo infall mass function also exhibits a strong dependence on $n_{\rm s}$ in a self-similar universe. However, if we focus only on the range of $\mu>10^{-3}$ covered by our simulations, the $n_s$ dependence aligns with what we observe in Figure~\ref{fig:cosmo_depend}, with a larger $n_s$ corresponding to a higher amplitude in the PMF while the slope barely varies. 

\revise{Finally, we emphasize that the EPS predictions here serve as a qualitative reference rather than a precise quantitative comparison to simulations. The theoretical mass function is not calibrated against simulation results using the halo mass definition adopted in this work, but instead directly follows the formulation of \citet{Lacey93}. Furthermore, the EPS framework implicitly assumes consistent mass definitions for halos and subhalos when calculating the peak mass ratio, whereas our analysis adopts different definitions for the two. These inconsistencies can result in deviations in both the amplitude and slope when comparing the predicted PMFs with simulation results. }



\end{document}